\documentstyle[12pt]{article}
\begin{document}
\def\beq{\begin{equation}}
\def\eeq{\end{equation}}
\def\bea{\begin{eqnarray}}
\def\eea{\end{eqnarray}}
\def\ve{\vert}
\def\vel{\left|}
\def\ver{\right|}
\def\nnb{\nonumber}
\def\ga{\left(}
\def\dr{\right)}
\def\aga{\left\{}
\def\adr{\right\}}
\def\rar{\rightarrow}
\def\nnb{\nonumber}
\def\la{\langle}
\def\ra{\rangle}
\def\lla{\left<}
\def\rra{\right>}
\def\ba{\begin{array}}
\def\ea{\end{array}}
\def\tep{$B \rar K \ell^+ \ell^-$}
\def\tepm{$B \rar K \mu^+ \mu^-$}
\def\tept{$B \rar K \tau^+ \tau^-$}
\def\ds{\displaystyle}



\newskip\humongous \humongous=0pt plus 1000pt minus 1000pt
\def\caja{\mathsurround=0pt}
\def\eqalign#1{\,\vcenter{\openup1\jot
\caja   \ialign{\strut \hfil$\displaystyle{##}$&$
\displaystyle{{}##}$\hfil\crcr#1\crcr}}\,}


\def\simlt{\stackrel{<}{{}_\sim}}
\def\simgt{\stackrel{>}{{}_\sim}}



\def\bos{\lower 0.5cm\hbox{{\vrule width 0pt height 1.2cm}}}
\def\boss{\lower 0.35cm\hbox{{\vrule width 0pt height 1.cm}}}
\def\aaa{\lower 0.cm\hbox{{\vrule width 0pt height .7cm}}}
\def\dol{\lower 0.4cm\hbox{{\vrule width 0pt height .5cm}}}


\title{ {\Large {\bf 
General analysis of lepton polarizations in rare 
$B \rar K^\ast \ell^+ \ell^-$ decay beyond the standard model } } }

\author{\vspace{1cm}\\
{\small T. M. Aliev \thanks
{e-mail: taliev@metu.edu.tr}\,\,,
M. K. \c{C}akmak\,\,,
M. Savc{\i} \thanks
{e-mail: savci@metu.edu.tr}} \\
{\small Physics Department, Middle East Technical University} \\
{\small 06531 Ankara, Turkey} }
\date{}

\begin{titlepage}
\maketitle
\thispagestyle{empty}

\begin{abstract}
The general analysis of lepton polarization asymmetries in rare 
$B \rar K^\ast \ell^+ \ell^-$ decay is investigated. Using the most general,
model independent effective Hamiltonian, the general expressions of the
longitudinal, normal and transversal polarization asymmetries for $\ell^-$
and $\ell^+$ and combinations of them are presented. The dependence of
lepton polarizations and their combinations on new Wilson coefficients are
studied in detail. Our analysis shows that the lepton polarization
asymmetries are very sensitive to the scalar and tensor type interactions,
which will be very useful in looking for new physics beyond the standard
model.
\end{abstract}

\end{titlepage}

\section{Introduction}
Rare $B$ meson decays, induced by flavor--changing neutral current (FCNC)
$b \rar s(d)$ transitions, are quite promising for establishing
new physics beyond the standard model (SM). In particular the flavor 
changing inclusive $b \rar s(d) \ell^+ \ell^-$ decay, which takes place in
the SM at loop level, is very sensitive to the gauge structure of the SM. 
Moreover $b \rar s(d) \ell^+ \ell^-$ decay is known to be very sensitive to
the various extensions of the SM. New physics effects manifest  
themselves in rare $B$
meson decays in two different ways, either through new contributions to the
Wilson coefficients existing in the SM or through
the new structures in the effective Hamiltonian which are absent in the SM.
Note that $b \rar s(d) \ell^+ \ell^-$ transition has been extensively
studied in framework of the SM and its various extensions
\cite{R1}--\cite{R15}. One of the efficient ways in establishing new physics
beyond the SM is the measurement of the lepton polarization
\cite{R15}--\cite{R24}. All previous studies for the lepton polarization
have been limited to SM and its minimal extensions, except the work
\cite{R19}. In \cite{R19} the analysis of the $\tau$ lepton polarization
for the inclusive $b \rar s \tau^+ \tau^-$ decay was presented in a model
independent way. In the same work, it was also shown that the investigation
of the $\tau$ lepton polarization can give unambiguous information about the
existence of the scalar and tensor type interactions. 

It is well known that the theoretical study of the inclusive decays is
rather easy but their experimental investigation is difficult. However for
the exclusive decays the situation is contrary to the inclusive case, i.e.,
their experimental detection is very easy but theoretical investigation
has its own drawbacks. This is due to the fact that for description of the exclusive
decay form factors, i.e., the matrix elements of the effective Hamiltonian
between initial and final meson states, are needed. This problem is related
to the nonperturbative sector of the QCD and and it can only be solved in
framework of the nonperturbative approaches. 

These matrix elements have been studied in framework of different
approaches, such as chiral theory \cite{R25}, three--point QCD sum rules
\cite{R26} and light cone QCD sum rules \cite{R27,R28}. In this work we will
use the weak decay form factors calculated using light cone QCD sum rules
method \cite{R27,R28}.

The aim of the present work is to present a rigorous study of the lepton
polarizations in the exclusive $B \rar K^\ast \ell^+\ell^-~(\ell =
\mu,~\tau)$ decay for a general form of the effective Hamiltonian including
tensor type interactions as well. 
In the present work we extend results of previous studies on lepton
polarization \cite{R20}--\cite{R23}
and then perform a
general analysis (in a model independent way in the sense without forcing
concrete values for the Wilson coefficients corresponding to any specific
model) including all possible form of interactions. 
Our analysis shows that the so--called tensor type interactions give
dominant contribution to the lepton polarization asymmetries.
The paper is organized as follows. In
section 2, using a general form of four--Fermi interaction  we derive the 
model independent expressions for the longitudinal,
transversal and normal polarizations of leptons. In section 3 we investigate
the dependence of the above--mentioned polarizations on the four--Fermi
interactions. We also present the combined analysis of the $\ell^-$ and
$\ell^+$ asymmetries and our results.

\section{Lepton polarizations}

We start this section by computing the lepton polarization asymmetries, using
the most general, model independent four--Fermi interactions. Following
\cite{R19}, we write the effective Hamiltonian for the $b \rar s \ell^+
\ell^-$ transition in terms of twelve model independent four--Fermi
interactions. 
\bea
\label{matel}
{\cal H}_{eff} &=& \frac{G\alpha}{\sqrt{2} \pi}
 V_{ts}V_{tb}^\ast
\Bigg\{ C_{SL} \, \bar s i \sigma_{\mu\nu} \frac{q^\nu}{q^2}\, L \,b  
\, \bar \ell \gamma^\mu \ell + C_{BR}\, \bar s i \sigma_{\mu\nu}
\frac{q^\nu}{q^2} \,R\, b \, \bar \ell \gamma^\mu \ell \nnb \\
&+&C_{LL}^{tot}\, \bar s_L \gamma_\mu b_L \,\bar \ell_L \gamma^\mu \ell_L +
C_{LR}^{tot} \,\bar s_L \gamma_\mu b_L \, \bar \ell_R \gamma^\mu \ell_R +  
C_{RL} \,\bar s_R \gamma_\mu b_R \,\bar \ell_L \gamma^\mu \ell_L \nnb \\
&+&C_{RR} \,\bar s_R \gamma_\mu b_R \, \bar \ell_R \gamma^\mu \ell_R +
C_{LRLR} \, \bar s_L b_R \,\bar \ell_L \ell_R +
C_{RLLR} \,\bar s_R b_L \,\bar \ell_L \ell_R \\
&+&C_{LRRL} \,\bar s_L b_R \,\bar \ell_R \ell_L +
C_{RLRL} \,\bar s_R b_L \,\bar \ell_R \ell_L+
C_T\, \bar s \sigma_{\mu\nu} b \,\bar \ell \sigma^{\mu\nu}\ell \nnb \\
&+&i C_{TE}\,\epsilon^{\mu\nu\alpha\beta} \bar s \sigma_{\mu\nu} b \,
\bar \ell \sigma_{\alpha\beta} \ell  \Bigg\}~, \nnb
\eea
where the chiral projection operators $L$ and $R$ in (\ref{matel}) are
defined as
\bea  
L = \frac{1-\gamma_5}{2} ~,~~~~~~ R = \frac{1+\gamma_5}{2}\nnb~,
\eea  
and $C_X$ are the coefficients of the four--Fermi interactions. The first
two of
these coefficients, $C_{SL}$ and $C_{BR}$, are the nonlocal Fermi
interactions which correspond to $-2 m_s C_7^{eff}$ and $-2 m_b C_7^{eff}$
in the SM, respectively. The following
four terms in this  
expression are the vector type interactions with
coefficients $C_{LL}$, $C_{LR}$, $C_{RL}$ and $C_{RR}$. Two of these
vector interactions containing $C_{LL}^{tot}$ and $C_{LR}^{tot}$ do already exist in the SM
in combinations of the form $(C_9^{eff}-C_{10})$ and $(C_9^{eff}+C_{10})$.
Therefore by writing 
\bea
C_{LL}^{tot} &=& C_9^{eff} - C_{10} + C_{LL}~, \nnb \\     
C_{LR}^{tot} &=& C_9^{eff} + C_{10} + C_{LR}~, \nnb
\eea
one concludes that $C_{LL}^{tot}$ and $C_{LR}^{tot}$ describe the
sum of the contributions from SM and the new physics. The terms with
coefficients $C_{LRLR}$, $C_{RLLR}$, $C_{LRRL}$ and $C_{RLRL}$ describe
the scalar type interactions. The remaining two terms leaded by the
coefficients $C_T$ and $C_{TE}$, obviously, describe the tensor type
interactions.    

Having the general form of four--Fermi interaction for the $b \rar s \ell^+
\ell^-$ transition, our next problem is calculation of the matrix element 
for the $B \rar K^\ast \ell^+ \ell^-$ decay. In other words, we need
the matrix elements 
\bea
\label{roll}
&&\lla K^\ast\vel \bar s \gamma_\mu (1 \pm \gamma_5) 
b \ver B \rra~,\nnb \\
&&\lla K^\ast \vel \bar s i\sigma_{\mu\nu} q^\nu  
(1 \pm \gamma_5) b \ver B \rra~, \nnb \\
&&\lla K^\ast \vel \bar s (1 \pm \gamma_5) b 
\ver B \rra~, \nnb \\
&&\lla K^\ast \vel \bar s \sigma_{\mu\nu} b
\ver B \rra~, \nnb
\eea
in order to calculate the decay amplitude for the $B \rar K^\ast \ell^+
\ell^-$ decay. These matrix elements are defined as follows:
\bea
\lefteqn{
\label{ilk}
\lla K^\ast(p_{K^\ast},\varepsilon) \vel \bar s \gamma_\mu 
(1 \pm \gamma_5) b \ver B(p_B) \rra =} \nnb \\
&&- \epsilon_{\mu\nu\lambda\sigma} \varepsilon^{\ast\nu} p_{K^\ast}^\lambda q^\sigma
\frac{2 V(q^2)}{m_B+m_{K^\ast}} \pm i \varepsilon_\mu^\ast (m_B+m_{K^\ast})   
A_1(q^2) \\
&&\mp i (p_B + p_{K^\ast})_\mu (\varepsilon^\ast q)
\frac{A_2(q^2)}{m_B+m_{K^\ast}}
\mp i q_\mu \frac{2 m_{K^\ast}}{q^2} (\varepsilon^\ast q)
\left[A_3(q^2)-A_0(q^2)\right]~,  \nnb \\  \nnb \\
\lefteqn{
\label{iki}
\lla K^\ast(p_{K^\ast},\varepsilon) \vel \bar s i \sigma_{\mu\nu} q^\nu
(1 \pm \gamma_5) b \ver B(p_B) \rra =} \nnb \\
&&4 \epsilon_{\mu\nu\lambda\sigma} 
\varepsilon^{\ast\nu} p_{K^\ast}^\lambda q^\sigma
T_1(q^2) \pm 2 i \left[ \varepsilon_\mu^\ast (m_B^2-m_{K^\ast}^2) -
(p_B + p_{K^\ast})_\mu (\varepsilon^\ast q) \right] T_2(q^2) \\
&&\pm 2 i (\varepsilon^\ast q) \left[ q_\mu -
(p_B + p_{K^\ast})_\mu \frac{q^2}{m_B^2-m_{K^\ast}^2} \right] 
T_3(q^2)~, \nnb \\  \nnb \\ 
\lefteqn{
\label{ucc}
\lla K^\ast(p_{K^\ast},\varepsilon) \vel \bar s \sigma_{\mu\nu} 
 b \ver B(p_B) \rra =} \nnb \\
&&i \epsilon_{\mu\nu\lambda\sigma}  \Bigg[ - 2 T_1(q^2)
{\varepsilon^\ast}^\lambda (p_B + p_{K^\ast})^\sigma +
\frac{2}{q^2} (m_B^2-m_{K^\ast}^2) {\varepsilon^\ast}^\lambda 
q^\sigma \\
&&- \frac{4}{q^2} \Bigg(T_1(q^2) - T_2(q^2) - \frac{q^2}{m_B^2-m_{K^\ast}^2} 
T_3(q^2) \Bigg) (\varepsilon^\ast q) p_{K^\ast}^\lambda q^\sigma \Bigg]~. \nnb 
\eea
where $q = p_B-p_{K^\ast}$ is the momentum transfer and $\varepsilon$ is the
polarization vector of $K^\ast$ meson. 
In order to ensure finiteness of (\ref{ilk}) at $q^2=0$, 
we assume that $A_3(q^2=0) = A_0(q^2=0)$ and $T_1(q^2=0) = T_2(q^2=0)$.
The matrix element $\lla K^\ast \vel \bar s (1 \pm \gamma_5 ) b \ver B \rra$
can be calculated from Eq. (\ref{ilk}) by 
contracting both sides of Eq. (\ref{ilk}) with $q^\mu$ and using equation of
motion. Neglecting the mass of the strange quark we get
\bea
\label{uc}
\lla K^\ast(p_{K^\ast},\varepsilon) \vel \bar s (1 \pm \gamma_5) b \ver
B(p_B) \rra =
\frac{1}{m_b} \Big[ \mp 2i m_{K^\ast} (\varepsilon^\ast q)
A_0(q^2)\Big]~.
\eea
In deriving Eq. (\ref{uc}) we have used the relation
(see \cite{R15,R26})
\bea
2 m_{K^\ast} A_3(q^2) = (m_B+m_{K^\ast}) A_1(q^2) -
(m_B-m_{K^\ast}) A_2(q^2)~. \nnb 
\eea
Taking into account Eqs. (\ref{matel}--\ref{uc}), the matrix element of the 
$B \rar K^\ast \ell^+ \ell^-$ decay can be written as 
\bea
\lefteqn{
\label{had}
{\cal M}(B\rightarrow K^\ast \ell^{+}\ell^{-}) =
\frac{G \alpha}{4 \sqrt{2} \pi} V_{tb} V_{ts}^\ast }\nnb \\
&&\times \Bigg\{
\bar \ell \gamma^\mu(1-\gamma_5) \ell \, \Big[
-2 A_1 \epsilon_{\mu\nu\lambda\sigma} \varepsilon^{\ast\nu}
p_{K^\ast}^\lambda q^\sigma
 -i B_1 \varepsilon_\mu^\ast
+ i B_2 (\varepsilon^\ast q) (p_B+p_{K^\ast})_\mu
+ i B_3 (\varepsilon^\ast q) q_\mu  \Big] \nnb \\
&&+ \bar \ell \gamma^\mu(1+\gamma_5) \ell \, \Big[
-2 C_1 \epsilon_{\mu\nu\lambda\sigma} \varepsilon^{\ast\nu}
p_{K^\ast}^\lambda q^\sigma
 -i D_1 \varepsilon_\mu^\ast    
+ i D_2 (\varepsilon^\ast q) (p_B+p_{K^\ast})_\mu
+ i D_3 (\varepsilon^\ast q) q_\mu  \Big] \nnb \\
&&+\bar \ell (1-\gamma_5) \ell \Big[ i B_4 (\varepsilon^\ast
q)\Big]
+ \bar \ell (1+\gamma_5) \ell \Big[ i B_5 (\varepsilon^\ast
q)\Big]  \nnb \\
&&+4 \bar \ell \sigma^{\mu\nu}  \ell \Big( i C_T \epsilon_{\mu\nu\lambda\sigma}
\Big) \Big[ -2 T_1 {\varepsilon^\ast}^\lambda (p_B+p_{K^\ast})^\sigma +
B_6 {\varepsilon^\ast}^\lambda q^\sigma -
B_7 (\varepsilon^\ast q) {p_{K^\ast}}^\lambda q^\sigma \Big] \nnb \\
&&+16 C_{TE} \bar \ell \sigma_{\mu\nu}  \ell \Big[ -2 T_1
{\varepsilon^\ast}^\mu (p_B+p_{K^\ast})^\nu  +B_6 {\varepsilon^\ast}^\mu q^\nu -
B_7 (\varepsilon^\ast q) {p_{K^\ast}}^\mu q^\nu
\Bigg\}~,
\eea
where
\bea
\label{as}
A_1 &=& (C_{LL}^{tot} + C_{RL}) \frac{V}{m_B+m_{K^\ast}} -
2 (C_{BR}+C_{SL}) \frac{T_1}{q^2} ~, \nnb \\
B_1 &=& (C_{LL}^{tot} - C_{RL}) (m_B+m_{K^\ast}) A_1 - 2
(C_{BR}-C_{SL}) (m_B^2-m_{K^\ast}^2)
\frac{T_2}{q^2} ~, \nnb \\
B_2 &=& \frac{C_{LL}^{tot} - C_{RL}}{m_B+m_{K^\ast}} A_2 - 2
(C_{BR}-C_{SL})
\frac{1}{q^2}  \left[ T_2 + \frac{q^2}{m_B^2-m_{K^\ast}^2} T_3 \right]~,
\nnb \\
B_3 &=& 2 (C_{LL}^{tot} - C_{RL}) m_{K^\ast} \frac{A_3-A_0}{q^2}+
2 (C_{BR}-C_{SL}) \frac{T_3}{q^2} ~, \nnb \\
C_1 &=& A_1 ( C_{LL}^{tot} \rar C_{LR}^{tot}~,~~C_{RL} \rar
C_{RR})~,\nnb \\
D_1 &=& B_1 ( C_{LL}^{tot} \rar C_{LR}^{tot}~,~~C_{RL} \rar
C_{RR})~,\nnb \\
D_2 &=& B_2 ( C_{LL}^{tot} \rar C_{LR}^{tot}~,~~C_{RL} \rar
C_{RR})~,\nnb \\
D_3 &=& B_3 ( C_{LL}^{tot} \rar C_{LR}^{tot}~,~~C_{RL} \rar
C_{RR})~,\nnb \\
B_4 &=& - 2 ( C_{LRRL} - C_{RLRL}) \frac{ m_{K^\ast}}{m_b} A_0 ~,\nnb \\
B_5 &=& - 2 ( C_{LRLR} - C_{RLLR}) \frac{m_{K^\ast}}{m_b} A_0 ~,\nnb \\
B_6 &=& 2 (m_B^2-m_{K^\ast}^2) \frac{T_1-T_2}{q^2} ~,\nnb \\
B_7 &=& \frac{4}{q^2} \left( T_1-T_2 - 
\frac{q^2}{m_B^2-m_{K^\ast}^2} T_3 \right)~.   
\eea
The form of Eq. (\ref{had}) reflects the fact that its difference from the SM case 
is due to the last four terms only, namely, scalar and tensor type interactions.
The next task to be considered is calculation of the final lepton
polarizations with the help of the matrix element for the 
$B \rar K^\ast \ell^+ \ell^-$ decay. For this purpose we we define the
following orthogonal unit vectors, $S_L^{-\mu}$ in the rest frame of
$\ell^-$ and $S_L^{+\mu}$ in the rest frame of $\ell^+$, for the
polarization of the leptons along the longitudinal ($L$), transversal ($T$)
and normal ($N$) directions:
\bea
\label{pol}
S_L^{-\mu} &\equiv& (0,\vec{e}_L^{\,-}) = 
\ga 0,\frac{\vec{p}_-}{\vel \vec{p}_- \ver} \dr~, \nnb \\
S_N^{-\mu} &\equiv& (0,\vec{e}_N^{\,-}) = 
\ga 0,\frac{\vec{p} \times \vec{p}_-}
{\vel \vec{p} \times \vec{p}_- \ver} \dr~, \nnb \\
S_T^{-\mu} &\equiv& (0,\vec{e}_T^{\,-}) = 
\ga 0, \vec{e}_N^{\,-} \times \vec{e}_L^{\,-} \dr~, \\
S_L^{+\mu} &\equiv& (0,\vec{e}_L^{\,+}) = 
\ga 0,\frac{\vec{p}_+}{\vel \vec{p}_+ \ver} \dr~, \nnb \\
S_N^{+\mu} &\equiv& (0,\vec{e}_N^{\,+}) = 
\ga 0,\frac{\vec{p} \times \vec{p}_+}
{\vel \vec{p} \times \vec{p}_+ \ver} \dr~, \nnb \\
S_T^{+\mu} &\equiv& (0,\vec{e}_T^{\,+}) = 
\ga 0, \vec{e}_N^{\,+} \times \vec{e}_L^{\,+} \dr~, \nnb
\eea
where $\vec{p}_\pm$ and $\vec{p}$ are the three momenta of $\ell^\pm$ and
$K^\ast$ meson in the center of mass (CM) frame of the $\ell^+ \ell^-$
system, respectively. The longitudinal unit vectors $S_L^-$ and $S_L^+$ are
boosted to CM frame of $\ell^+ \ell^-$ by Lorentz transformation,
\bea
\label{bs}
S^{-\mu}_{L,\, CM} &=& \ga \frac{\vel \vec{p}_- \ver}{m_\ell}, 
\frac{E_\ell \,\vec{p}_-}{m_\ell \vel \vec{p}_- \ver} \dr~, \nnb \\
S^{+\mu}_{L,\, CM} &=& \ga \frac{\vel \vec{p}_- \ver}{m_\ell}, 
- \frac{E_\ell \, \vec{p}_-}{m_\ell \vel \vec{p}_- \ver} \dr~,
\eea
while vectors of perpendicular directions are not changed by boost.

The differential decay rate of the $B \rar K^\ast \ell^+ \ell^-$ decay for
any spin direction $\vec{n}^{\,(\pm)}$ of the $\ell^{(\pm)}$, where 
$\vec{n}^{\,(\pm)}$ is the unit vector in the $\ell^{(\pm)}$ rest frame,
can be written in the following form
\bea
\label{ddr}
\frac{d\Gamma(\vec{n}^{(\pm)})}{dq^2} = \frac{1}{2} 
\ga \frac{d\Gamma}{dq^2}\dr_0  
\Bigg[ 1 + \Bigg( P_L^{(\pm)} \vec{e}_L^{\,(\pm)} + P_N^{(\pm)}
\vec{e}_N^{\,(\pm)} + P_T^{(\pm)} \vec{e}_T^{\,(\pm)} \Bigg) \cdot
\vec{n}^{(\pm)} \Bigg]~,
\eea
where the superscripts $^+$ and $^-$ correspond to $\ell^+$ and $\ell^-$
cases, the subscript $_0$ corresponds to the unpolarized decay
rate, whose explicit form will be presented below and   
$P_L$, $P_N$ and $P_T$ represent the longitudinal, normal and transversal
polarizations, respectively.
The explicit form of the unpolarized decay rate in Eq. (\ref{ddr}) is

\bea
\label{unp}
\lefteqn{
\ga \frac{d \Gamma}{dq^2}\dr_0 = \frac{G^2 \alpha^2}{2^{14} \pi^5 m_B} 
\vel V_{tb} V_{ts}^\ast \ver^2 \lambda^{1/2} v} \nnb \\
&\times& \Bigg\{
\frac{32}{3} m_B^4 \lambda \Big[(m_B^2 s - m_\ell^2)
\ga \vel  A_1 \ver^2 + \vel  C_1 \ver^2 \dr + 6 m_\ell^2 \, \mbox{\rm Re} 
(A_1 C_1^\ast)\Big] \nnb \\
&+& 96 m_\ell^2 \, \mbox{\rm Re} (B_1 D_1^\ast)-
\frac{4}{r} m_B^2 m_\ell \lambda \,
\mbox{\rm Re} [(B_1 - D_1) (B_4^\ast - B_5^\ast)] \nnb \\
&+&\frac{8}{r} m_B^2 m_\ell^2 \lambda \, \Big(
\mbox{\rm Re} [B_1 (- B_3^\ast + D_2^\ast + D_3^\ast)] +
\mbox{\rm Re} [D_1 (B_2^\ast + B_3^\ast - D_3^\ast)] -
\mbox{\rm Re}(B_4 B_5^\ast) \Big] \Big)~~~~~~~~ \nnb \\
&+&\frac{4}{r} m_B^4 m_\ell (1-r) \lambda \,
\Big(\mbox{\rm Re} [(B_2 - D_2) (B_4^\ast - B_5^\ast)]
+2 m_\ell \, \mbox{\rm Re} [(B_2 - D_2) (B_3^\ast - D_3^\ast)]
\Big) \nnb \\
&-& \frac{8}{r}m_B^4 m_\ell^2 \lambda (2+2 r-s)\, \mbox{\rm Re} (B_2 D_2^\ast)
+\frac{4}{r} m_B^4 m_\ell s \lambda \,
\mbox{\rm Re} [(B_3 - D_3) (B_4^\ast - B_5^\ast)] \nnb \\
&+&\frac{4}{r} m_B^4 m_\ell^2 s \lambda \,
\vel B_3 - D_3\ver^2
+\frac{2}{r} m_B^2 (m_B^2 s-2 m_\ell^2) \lambda \,
\ga \vel B_4 \ver^2 + \vel B_5 \ver^2 \dr \nnb \\
&-&\frac{8}{3rs} m_B^2 \lambda \,
\Big[m_\ell^2 (2-2 r+s)+m_B^2 s (1-r-s) \Big]
\Big[\mbox{\rm Re}(B_1 B_2^\ast) + \mbox{\rm Re}(D_1 D_2^\ast)\Big] \nnb \\ 
&+&\frac{4}{3rs}\,
\Big[2 m_\ell^2 (\lambda-6 rs)+m_B^2 s (\lambda+12 rs) \Big]
\ga \vel B_1 \ver^2 + \vel D_1 \ver^2 \dr \nnb \\
&+&\frac{4}{3rs} m_B^4 \lambda\,
\Big( m_B^2 s \lambda + m_\ell^2 [ 2 \lambda + 3 s (2+2 r - s) ] \Big)
\ga \vel B_2 \ver^2 + \vel D_2 \ver^2 \dr \nnb \\
&+&\frac{32}{r} m_B^6 m_\ell \lambda^2 \,
\mbox{\rm Re} [(B_2 + D_2)(B_7 C_{TE})^\ast]  \\ 
&-& \frac{32}{r} m_B^4 m_\ell \lambda (1-r-s) \Big(
\mbox{\rm Re} [(B_1 + D_1)(B_7 C_{TE})^\ast] + 
2\, \mbox{\rm Re} [(B_2 + D_2)(B_6 C_{TE})^\ast] \Big) \nnb \\
&+&\frac{64}{r} (\lambda+12 rs) m_B^2 m_\ell \, 
\mbox{\rm Re} [(B_1 + D_1)(B_6 C_{TE})^\ast] \nnb \\
&+&\frac{256}{3rs} \vel t_1 \ver^2 \vel C_T \ver^2 m_B^2 
\Big( 4 m_\ell^2 \, [ \lambda ( 8r -s) - 12 r s (2 +2r -s) ] \nnb \\
&+& m_B^2 s \, [\lambda (16r+s)+12 r s (2 +2r -s) ] \Big) \nnb \\
&+&\frac{1024}{3rs} \vel t_1 \ver^2 \vel C_{TE} \ver^2 m_B^2 
\Big( 8 m_\ell^2 \, [ \lambda ( 4r+s) + 12 r s (2 +2r -s) ] \nnb \\ 
&+& m_B^2 s \, [\lambda (16r+s)+12 r s (2 +2r -s) ] \Big) \nnb \\
&-& \frac{128}{r} m_B^2 m_\ell \, [\lambda + 12 r (1-r) ] 
\,\mbox{\rm Re} [(B_1 + D_1)(t_1 C_{TE})^\ast] \nnb \\
&+&\frac{128}{r} m_B^4 m_\ell \lambda (1+3r-s)  
\,\mbox{\rm Re} [(B_2 + D_2)(t_1 C_{TE})^\ast]
+ 512 m_B^4 m_\ell \lambda \,
\mbox{\rm Re} [(A_1 + C_1)(t_1 C_T)^\ast] \nnb \\
&+&\frac{16}{3r} m_B^2 \Bigg( 4 (m_B^2 s + 8 m_\ell^2) \vel C_{TE} \ver^2 
+ m_B^2 s v^2 \vel C_T \ver^2 \Bigg)
\times \Bigg( 4 (\lambda+12 r s) \vel B_6 \ver^2 \nnb \\
&+& m_B^4 \lambda^2 \vel B_7 \ver^2 
- 4 m_B^2 (1-r-s) \lambda \,\mbox{\rm Re} (B_6 B_7^\ast)
- 16 \, [\lambda + 12 r (1-r) ]\,\mbox{\rm Re} (t_1 B_6^\ast) \nnb \\
&+& 8 m_B^2 (1+3r-s) \lambda \,\mbox{\rm Re} (t_1 B_7^\ast)\Bigg)\nnb
\Bigg\}~,
\eea
where $s=q^2/m_B^2$, $r=m_{K^\ast}^2/m_B^2$ and $v=\sqrt{1-\ds{\frac{4
m_\ell^2}{q^2}}}$ is the lepton velocity.

The polarizations $P_L$, $P_N$ and $P_T$ are defined as:
\bea
P_i^{(\pm)}(q^2) = \frac{\ds{\frac{d \Gamma}{dq^2}
                   (\vec{n}^{(\pm)}=\vec{e}_i^{\,(\pm)}) -
                   \frac{d \Gamma}{dq^2}
                   (\vec{n}^{(\pm)}=-\vec{e}_i^{\,(\pm)})}}
              {\ds{\frac{d \Gamma}{dq^2}
                   (\vec{n}^{(\pm)}=\vec{e}_i^{\,(\pm)}) +
                  \frac{d \Gamma}{dq^2}
                  (\vec{n}^{(\pm)}=-\vec{e}_i^{\,(\pm)})}}~, \nnb 
\eea
where $P^{(\pm)}$ represents the charged lepton $\ell^{(\pm)}$ 
polarization asymmetry for $i=L,~N,~T$, i.e., $P_L$ and $P_T$ are the
longitudinal and transversal asymmetries in the decay plane, respectively,
and $P_N$ is the normal component to both of them. With respect to the direction
of the lepton polarization, $P_L$ and $P_T$ are $P$--odd, $T$--even, while
$P_N$ is $P$--even, $T$--odd and $CP$--odd. 
After lengthy calculations for the longitudinal polarization of the
$\ell^{(\pm)}$, we get
\bea
\label{plm}
P_L^-&=& \frac{4}{\Delta} m_B^2 v \Bigg\{ 
\frac{1}{3 r} \lambda^2 m_B^4 \Big[ \vel B_2 \ver^2 - \vel D_2 \ver^2\Big] +
\frac{1}{r} \lambda  m_\ell \, 
\mbox{\rm Re} [(B_1 - D_1) (B_4^\ast + B_5^\ast)] \nnb \\
&-& \frac{1}{r} \lambda m_B^2 m_\ell (1-r) \, 
\mbox{\rm Re} [(B_2 - D_2) (B_4^\ast + B_5^\ast)]+
\frac{8}{3} \lambda m_B^4 s \Big[ \vel A_1 \ver^2 - \vel C_1 \ver^2\Big]
\nnb \\
&-&\frac{1}{2r} \lambda m_B^2 s 
\Big[ \vel B_4 \ver^2 - \vel B_5 \ver^2\Big]-
\frac{1}{r} \lambda m_B^2 m_\ell s \, 
\mbox{\rm Re} [(B_3 - D_3) (B_4^\ast + B_5^\ast)] \nnb \\
&-&\frac{2}{3 r} \lambda m_B^2 (1-r-s)
\Big[ \mbox{\rm Re}(B_1 B_2^\ast) - \mbox{\rm Re}(D_1 D_2^\ast)\Big]
+\frac{1}{3 r} (\lambda + 12 r s) 
\Big[ \vel B_1 \ver^2 - \vel D_1 \ver^2\Big] \nnb \\
&+&\frac{256}{3} \lambda m_B^2 m_\ell 
\,\Big( \mbox{\rm Re} [A_1^\ast (C_{T} + C_{TE})t_1] -
\mbox{\rm Re} [C_1^\ast (C_T- C_{TE})t_1] \Big) \nnb \\
&+&\frac{4}{3r} \lambda^2 m_B^4 m_\ell \Big(
\mbox{\rm Re} [B_2^\ast (C_T+4 C_{TE}) B_7]
+ \mbox{\rm Re} [D_2^\ast (C_T-4 C_{TE}) B_7] \Big)  \nnb \\
&-&\frac{8}{3r} \lambda m_B^2 m_\ell (1-r-s) \Big(
\mbox{\rm Re} [B_2^\ast (C_T+4 C_{TE}) B_6]
+\mbox{\rm Re} [D_2^\ast (C_T-4 C_{TE}) B_6] \Big) \nnb \\
&-&\frac{4}{3r} \lambda m_B^2 m_\ell (1-r-s) \Big(
\mbox{\rm Re} [B_1^\ast (C_T+4 C_{TE}) B_7]
+ \mbox{\rm Re} [D_1^\ast (C_T-4 C_{TE}) B_7]  \Big) \nnb \\
&+&\frac{8}{3r} (\lambda+12 r s)  m_\ell \Big(
\,\mbox{\rm Re} [B_1^\ast (C_T+4 C_{TE}) B_6]
+\mbox{\rm Re} [D_1^\ast (C_T-4 C_{TE}) B_6] \Big) \nnb \\
&-&\frac{16}{3r}  m_\ell [\lambda+12 r (1-r)] \Big(
\mbox{\rm Re} [B_1^\ast (C_T+4 C_{TE}) t_1]
+\mbox{\rm Re} [D_1^\ast (C_T-4 C_{TE}) t_1] \Big)\\
&+&\frac{16}{3r} \lambda m_B^2 m_\ell (1+3 r - s) \Big(             
\mbox{\rm Re} [B_2^\ast (C_T+4 C_{TE}) t_1]
+\mbox{\rm Re} [D_2^\ast (C_T-4 C_{TE}) t_1] \Big) \nnb \\
&+&\frac{16}{3r} \lambda^2 m_B^6 s 
\vel B_7 \ver^2 \mbox{\rm Re} (C_T C_{TE}^\ast) \nnb \\
&+&\frac{64}{3r} (\lambda + 12 r s) m_B^2 s     
\vel B_6 \ver^2 \mbox{\rm Re} (C_T C_{TE}^\ast) \nnb \\
&-&\frac{64}{3r} \lambda m_B^4 s (1-r-s)    
\,\mbox{\rm Re} (B_6 B_7^\ast) \mbox{\rm Re} (C_T C_{TE}^\ast) \nnb \\
&+&\frac{128}{3r} \lambda m_B^4 s (1+3 r-s) 
\,\mbox{\rm Re} (B_7 t_1^\ast) \mbox{\rm Re} (C_T C_{TE}^\ast) \nnb \\
&-&\frac{256}{3r} m_B^2 s [\lambda + 12 r (1-r)]  
\,\mbox{\rm Re} (B_6 t_1^\ast) \mbox{\rm Re} (C_T C_{TE}^\ast) \nnb \\
&+&\frac{256}{3r} m_B^2 
[ \lambda (4 r + s) + 12 r (1-r)^2 ] \vel t_1 \ver^2
\,\mbox{\rm Re} (C_T C_{TE}^\ast)
\Bigg\}~,\nnb          
\eea
\bea
\label{plp}
P_L^+&=& \frac{4}{\Delta} m_B^2 v \Bigg\{ 
- \frac{1}{3 r} \lambda^2 m_B^4 \Big[ \vel B_2 \ver^2 - \vel D_2 \ver^2\Big] +
\frac{1}{r} \lambda  m_\ell \, 
\mbox{\rm Re} [(B_1 - D_1) (B_4^\ast + B_5^\ast)] \nnb \\
&-& \frac{1}{r} \lambda m_B^2 m_\ell (1-r) \, 
\mbox{\rm Re} [(B_2 - D_2) (B_4^\ast + B_5^\ast)] -
\frac{8}{3} \lambda m_B^4 s \Big[ \vel A_1 \ver^2 - \vel C_1 \ver^2\Big]
\nnb \\
&-&\frac{1}{2r} \lambda m_B^2 s 
\Big[ \vel B_4 \ver^2 - \vel B_5 \ver^2\Big]-
\frac{1}{r} \lambda m_B^2 m_\ell s \, 
\mbox{\rm Re} [(B_3 - D_3) (B_4^\ast + B_5^\ast)] \nnb \\
&+&\frac{2}{3 r} \lambda m_B^2 (1-r-s)
\Big[ \mbox{\rm Re}(B_1 B_2^\ast) - \mbox{\rm Re}(D_1 D_2^\ast)\Big]
- \frac{1}{3 r} (\lambda + 12 r s) 
\Big[ \vel B_1 \ver^2 - \vel D_1 \ver^2\Big] \nnb \\
&-&\frac{256}{3} \lambda m_B^2 m_\ell 
\,\Big( \mbox{\rm Re} [A_1^\ast (C_{T} - C_{TE})t_1] -
\mbox{\rm Re} [C_1^\ast (C_T + C_{TE})t_1] \Big) \nnb \\
&+&\frac{4}{3r} \lambda^2 m_B^4 m_\ell \Big(
\mbox{\rm Re} [B_2^\ast (C_T-4 C_{TE}) B_7]
+ \mbox{\rm Re} [D_2^\ast (C_T+4 C_{TE}) B_7] \Big)  \nnb \\
&-&\frac{8}{3r} \lambda m_B^2 m_\ell (1-r-s) \Big(
\mbox{\rm Re} [B_2^\ast (C_T-4 C_{TE}) B_6]
+\mbox{\rm Re} [D_2^\ast (C_T+4 C_{TE}) B_6] \Big) \nnb \\
&-&\frac{4}{3r} \lambda m_B^2 m_\ell (1-r-s) \Big(
\mbox{\rm Re} [B_1^\ast (C_T-4 C_{TE}) B_7]
+ \mbox{\rm Re} [D_1^\ast (C_T+4 C_{TE}) B_7]  \Big) \nnb \\
&+&\frac{8}{3r} (\lambda+12 r s)  m_\ell \Big(
\,\mbox{\rm Re} [B_1^\ast (C_T-4 C_{TE}) B_6]
+\mbox{\rm Re} [D_1^\ast (C_T+4 C_{TE}) B_6] \Big) \nnb \\
&-&\frac{16}{3r}  m_\ell [\lambda+12 r (1-r)] \Big(
\mbox{\rm Re} [B_1^\ast (C_T-4 C_{TE}) t_1]
+\mbox{\rm Re} [D_1^\ast (C_T+4 C_{TE}) t_1] \Big) \\
&+&\frac{16}{3r} \lambda m_B^2 m_\ell (1+3 r - s) \Big(             
\mbox{\rm Re} [B_2^\ast (C_T-4 C_{TE}) t_1]
+\mbox{\rm Re} [D_2^\ast (C_T+4 C_{TE}) t_1] \Big) \nnb \\
&+&\frac{16}{3r} \lambda^2 m_B^6 s 
\vel B_7 \ver^2 \mbox{\rm Re} (C_T C_{TE}^\ast) \nnb \\
&+&\frac{64}{3r} (\lambda + 12 r s) m_B^2 s     
\vel B_6 \ver^2 \mbox{\rm Re} (C_T C_{TE}^\ast) \nnb \\
&-&\frac{64}{3r} \lambda m_B^4 s (1-r-s)    
\,\mbox{\rm Re} (B_6 B_7^\ast) \mbox{\rm Re} (C_T C_{TE}^\ast) \nnb \\
&+&\frac{128}{3r} \lambda m_B^4 s (1+3 r-s) 
\,\mbox{\rm Re} (B_7 t_1^\ast) \mbox{\rm Re} (C_T C_{TE}^\ast) \nnb \\
&-&\frac{256}{3r} m_B^2 s [\lambda + 12 r (1-r)]  
\,\mbox{\rm Re} (B_6 t_1^\ast) \mbox{\rm Re} (C_T C_{TE}^\ast) \nnb \\
&+&\frac{256}{3r} m_B^2 
[ \lambda (4 r + s) + 12 r (1-r)^2 ] \vel t_1 \ver^2
\,\mbox{\rm Re} (C_T C_{TE}^\ast)
\Bigg\}~, \nnb          
\eea
where $\Delta$ is the term inside curly brackets of Eq. (\ref{unp}). From 
Eqs. (\ref{plm}) and (\ref{plp}) we observe that the terms containing "pure"
SM contribution, i.e., the terms containing $C_{BR},~C_{SL},~C_{LL}^{tot}$ and
$C_{LR}^{tot}$ are the same for both lepton and antilepton but with opposite sign.
However for the terms containing new physics effects this does not hold. In
other words, such terms may have same or different signs for lepton and
antilepton. In due course this difference  
can be a useful tool for looking new physics effects.

Further calculations lead the following expressions for the transverse
polarization $P_T^{(\pm)}$:
\bea
\label{ptm}
P_T^-&=& \frac{\pi}{\Delta} m_B \sqrt{s \lambda} \Bigg\{ 
-8 m_B^2 m_\ell  \, \mbox{\rm Re} [(A_1 + C_1) (B_1^\ast + D_1^\ast)] \nnb \\
&+& \frac{1}{r} m_B^2 m_\ell (1+3 r - s)  \, 
\Big[ \mbox{\rm Re}(B_1 D_2^\ast) -  \mbox{\rm Re}(B_2 D_1^\ast)\Big] \nnb \\
&+&\frac{1}{r s} m_\ell (1- r - s)
\Big[ \vel B_1 \ver^2 - \vel D_1 \ver^2\Big] \nnb \\
&+&\frac{2}{r s} m_\ell^2 (1- r - s)
\Big[ \mbox{\rm Re}(B_1 B_5^\ast) - \mbox{\rm Re}(D_1 B_4^\ast)\Big] \nnb \\
&-&\frac{1}{r} m_B^2 m_\ell (1- r - s) \,
\mbox{\rm Re} [(B_1 + D_1) (B_3^\ast - D_3^\ast)] \nnb \\
&-&\frac{2}{r s} m_B^2 m_\ell^2 \lambda 
\Big[  \mbox{\rm Re}(B_2 B_5^\ast) -  \mbox{\rm Re}(D_2 B_4^\ast)\Big] \nnb \\ 
&+&\frac{1}{r s} m_B^4 m_\ell(1-r) \lambda
\Big[ \vel B_2 \ver^2 - \vel D_2 \ver^2\Big]
+ \frac{1}{r} m_B^4 m_\ell \lambda  \,
\mbox{\rm Re} [(B_2 + D_2) (B_3^\ast - D_3^\ast)] \nnb \\
&-&\frac{1}{r s} m_B^2 m_\ell [\lambda + (1-r-s) ( 1-r)]
\Big[\mbox{\rm Re}(B_1 B_2^\ast) - \mbox{\rm Re}(D_1 D_2^\ast)\Big] \nnb \\
&+&\frac{1}{r s} (1-r-s)(2 m_\ell^2  - m_B^2 s )
\Big[\mbox{\rm Re}(B_1 B_4^\ast)-\mbox{\rm Re}(D_1 B_5^\ast)\Big] \nnb \\
&+&\frac{1}{r s} m_B^2 \lambda (2 m_\ell^2  - m_B^2 s )
\Big[\mbox{\rm Re}(D_2 B_5^\ast) - \mbox{\rm Re}(B_2 B_4^\ast)\Big] \nnb \\
&-&\frac{16}{r s} \lambda m_B^2 m_\ell^2 
\,\mbox{\rm Re}[(B_1-D_1) (B_7 C_{TE})^\ast] \nnb \\   
&+&\frac{16}{r s} \lambda m_B^4 m_\ell^2 (1-r)    
\,\mbox{\rm Re}[(B_2-D_2) (B_7 C_{TE})^\ast] \nnb \\
&+&\frac{8}{r} \lambda m_B^4 m_\ell 
\,\mbox{\rm Re}[(B_4-B_5) (B_7 C_{TE})^\ast] \nnb \\
&+&\frac{16}{r} \lambda m_B^4 m_\ell^2       
\,\mbox{\rm Re}[(B_3-D_3) (B_7 C_{TE})^\ast] \nnb \\
&+&\frac{32}{r s} m_\ell^2 (1-r-s)       
\,\mbox{\rm Re}[(B_1-D_1) (B_6 C_{TE})^\ast] \\
&-&\frac{32}{r s} m_B^2 m_\ell^2 (1-r) (1-r-s)       
\,\mbox{\rm Re}[(B_2-D_2) (B_6 C_{TE})^\ast] \nnb \\
&-&\frac{16}{r} m_B^2 m_\ell  (1-r-s)      
\,\mbox{\rm Re}[(B_4-B_5) (B_6 C_{TE})^\ast] \nnb \\
&-&\frac{32}{r} m_B^2 m_\ell^2  (1-r-s)      
\,\mbox{\rm Re}[(B_3-D_3) (B_6 C_{TE})^\ast] \nnb \\
&-& 16 m_B^2  \Big(
4 m_\ell^2 \, \mbox{\rm Re}[A_1^\ast (C_T+2 C_{TE}) B_6]
- m_B^2 s \, \mbox{\rm Re}[A_1^\ast (C_T-2 C_{TE}) B_6] \Big)\nnb \\
&+& 16 m_B^2  \Big(
4 m_\ell^2 \, \mbox{\rm Re}[C_1^\ast (C_T-2 C_{TE}) B_6]  
- m_B^2 s \, \mbox{\rm Re}[C_1^\ast (C_T+2 C_{TE}) B_6] \Big)\nnb \\
&+& \frac{32}{s} m_B^2 (1-r) \Big(
4 m_\ell^2 \, \mbox{\rm Re}[A_1^\ast (C_T+2 C_{TE}) t_1]
- m_B^2 s \, \mbox{\rm Re}[A_1^\ast (C_T-2 C_{TE}) t_1] \Big)\nnb \\
&-& \frac{32}{s} m_B^2 (1-r) \Big(
4 m_\ell^2 \, \mbox{\rm Re}[C_1^\ast (C_T-2 C_{TE}) t_1]
- m_B^2 s \, \mbox{\rm Re}[C_1^\ast (C_T+2 C_{TE}) t_1] \Big)\nnb \\
&+&\frac{64}{r s} m_B^2 m_\ell^2 (1-r) (1+3 r-s)       
\,\mbox{\rm Re}[(B_2- D_2) (t_1 C_{TE})^\ast] \nnb \\
&+&\frac{64}{r} m_B^2 m_\ell^2  (1+3 r-s)       
\,\mbox{\rm Re}[(B_3-D_3) (t_1 C_{TE})^\ast] \nnb \\
&+&\frac{32}{r} m_B^2 m_\ell  (1+3 r-s)
\,\mbox{\rm Re}[(B_4-B_5) (t_1 C_{TE})^\ast] \nnb \\
&+&\frac{64}{r s} [m_B^2 r s - m_\ell^2 (1+7 r -s)] 
\,\mbox{\rm Re}[(B_1-D_1) (t_1 C_{TE})^\ast] \nnb \\
&-& \frac{32}{s} (4 m_\ell^2 + m_B^2 s)
\,\mbox{\rm Re}[(B_1+D_1) (t_1 C_T)^\ast] \nnb \\
&-&2048  m_B^2 m_\ell  
\,\mbox{\rm Re}[(C_T t_1)(B_6 C_{TE})^\ast] \nnb \\
&+&\frac{4096}{s} m_B^2 m_\ell (1-r) \vel t_1 \ver^2 
\, \mbox{\rm Re}(C_T C_{TE}^\ast)
\Bigg\}~, \nnb
\eea
\bea
\label{ptp}
P_T^+&=& \frac{\pi}{\Delta} m_B \sqrt{s \lambda} \Bigg\{ 
-8 m_B^2 m_\ell  \, \mbox{\rm Re} [(A_1 + C_1) (B_1^\ast + D_1^\ast)] \nnb \\
&-& \frac{1}{r} m_B^2 m_\ell (1+3 r - s)  \, 
\Big[ \mbox{\rm Re}(B_1 D_2^\ast) -  \mbox{\rm Re}(B_2 D_1^\ast)\Big] \nnb \\
&-&\frac{1}{r s} m_\ell (1- r - s)
\Big[ \vel B_1 \ver^2 - \vel D_1 \ver^2\Big] \nnb \\
&+&\frac{1}{r s} (2 m_\ell^2  - m_B^2 s ) (1- r - s)
\Big[ \mbox{\rm Re}(B_1 B_5^\ast) - \mbox{\rm Re}(D_1 B_4^\ast)\Big] \nnb \\
&+&\frac{1}{r} m_B^2 m_\ell (1- r - s) \,
\mbox{\rm Re} [(B_1 + D_1) (B_3^\ast - D_3^\ast)] \nnb \\
&-&\frac{1}{r s} m_B^2 \lambda (2 m_\ell^2  - m_B^2 s ) 
\Big[  \mbox{\rm Re}(B_2 B_5^\ast) -  \mbox{\rm Re}(D_2 B_4^\ast)\Big] \nnb \\ 
&-&\frac{1}{r s} m_B^4 m_\ell(1-r) \lambda
\Big[ \vel B_2 \ver^2 - \vel D_2 \ver^2\Big]
- \frac{1}{r} m_B^4 m_\ell \lambda  \,
\mbox{\rm Re} [(B_2 + D_2) (B_3^\ast - D_3^\ast)] \nnb \\
&+&\frac{1}{r s} m_B^2 m_\ell [\lambda + (1-r-s) ( 1-r)]
\Big[\mbox{\rm Re}(B_1 B_2^\ast) - \mbox{\rm Re}(D_1 D_2^\ast)\Big] \nnb \\
&+&\frac{2}{r s}  m_\ell^2 (1-r-s)
\Big[\mbox{\rm Re}(B_1 B_4^\ast)-\mbox{\rm Re}(D_1 B_5^\ast)\Big] \nnb \\
&+&\frac{2}{r s} m_B^2  m_\ell^2 \lambda
\Big[\mbox{\rm Re}(D_2 B_5^\ast) - \mbox{\rm Re}(B_2 B_4^\ast)\Big] \nnb \\
&+&\frac{16}{r s} \lambda m_B^2 m_\ell^2 
\,\mbox{\rm Re}[(B_1-D_1) (B_7 C_{TE})^\ast] \nnb \\   
&-&\frac{16}{r s} \lambda m_B^4 m_\ell^2 (1-r)    
\,\mbox{\rm Re}[(B_2-D_2) (B_7 C_{TE})^\ast] \nnb \\
&-&\frac{8}{r} \lambda m_B^4 m_\ell 
\,\mbox{\rm Re}[(B_4-B_5) (B_7 C_{TE})^\ast] \nnb \\
&-&\frac{16}{r} \lambda m_B^4 m_\ell^2       
\,\mbox{\rm Re}[(B_3-D_3) (B_7 C_{TE})^\ast] \nnb \\
&-&\frac{32}{r s} m_\ell^2 (1-r-s)       
\,\mbox{\rm Re}[(B_1-D_1) (B_6 C_{TE})^\ast] \\
&+&\frac{32}{r s} m_B^2 m_\ell^2 (1-r) (1-r-s)       
\,\mbox{\rm Re}[(B_2-D_2) (B_6 C_{TE})^\ast] \nnb \\
&+&\frac{16}{r} m_B^2 m_\ell  (1-r-s)      
\,\mbox{\rm Re}[(B_4-B_5) (B_6 C_{TE})^\ast] \nnb \\
&+&\frac{32}{r} m_B^2 m_\ell^2  (1-r-s)      
\,\mbox{\rm Re}[(B_3-D_3) (B_6 C_{TE})^\ast] \nnb \\
&+& 16 m_B^2  \Big(
4 m_\ell^2 \, \mbox{\rm Re}[A_1^\ast (C_T-2 C_{TE}) B_6]
- m_B^2 s \, \mbox{\rm Re}[A_1^\ast (C_T+2 C_{TE}) B_6] \Big)\nnb \\
&-& 16 m_B^2  \Big(
4 m_\ell^2 \, \mbox{\rm Re}[C_1^\ast (C_T+2 C_{TE}) B_6]  
- m_B^2 s \, \mbox{\rm Re}[C_1^\ast (C_T-2 C_{TE}) B_6] \Big)\nnb \\
&-& \frac{32}{s} m_B^2 (1-r) \Big(
4 m_\ell^2 \, \mbox{\rm Re}[A_1^\ast (C_T-2 C_{TE}) t_1]
- m_B^2 s \, \mbox{\rm Re}[A_1^\ast (C_T+2 C_{TE}) t_1] \Big)\nnb \\
&+& \frac{32}{s} m_B^2 (1-r) \Big(
4 m_\ell^2 \, \mbox{\rm Re}[C_1^\ast (C_T+2 C_{TE}) t_1]
- m_B^2 s \, \mbox{\rm Re}[C_1^\ast (C_T-2 C_{TE}) t_1] \Big)\nnb \\
&-&\frac{64}{r s} m_B^2 m_\ell^2 (1-r) (1+3 r-s)       
\,\mbox{\rm Re}[(B_2- D_2) (t_1 C_{TE})^\ast] \nnb \\
&-&\frac{64}{r} m_B^2 m_\ell^2  (1+3 r-s)       
\,\mbox{\rm Re}[(B_3-D_3) (t_1 C_{TE})^\ast] \nnb \\
&-&\frac{32}{r} m_B^2 m_\ell  (1+3 r-s)
\,\mbox{\rm Re}[(B_4-B_5) (t_1 C_{TE})^\ast] \nnb \\
&-&\frac{64}{r s} [m_B^2 r s - m_\ell^2 (1+7 r -s)] 
\,\mbox{\rm Re}[(B_1-D_1) (t_1 C_{TE})^\ast] \nnb \\
&-& \frac{32}{s} (4 m_\ell^2 + m_B^2 s)
\,\mbox{\rm Re}[(B_1+D_1) (t_1 C_T)^\ast] \nnb \\
&-&2048  m_B^2 m_\ell  
\,\mbox{\rm Re}[(C_T t_1)(B_6 C_{TE})^\ast] \nnb \\
&+&\frac{4096}{s} m_B^2 m_\ell (1-r) \vel t_1 \ver^2 
\, \mbox{\rm Re}(C_T C_{TE}^\ast)
\Bigg\}~. \nnb
\eea

Finally for normal asymmetries we get

\bea
\label{pnm}
P_N^-&=& \frac{1}{\Delta} \pi v m_B^3 \sqrt{s \lambda} \Bigg\{
8 m_\ell \, \mbox{\rm Im}[(B_1^\ast C_1) + (A_1^\ast D_1)] \nnb \\
&-& \frac{1}{r} m_B^2 \lambda 
\,\mbox{\rm Im}[(B_2^\ast B_4) + (D_2^\ast B_5)] \nnb \\
&+& \frac{1}{r} m_B^2 m_\ell \lambda 
\,\mbox{\rm Im}[(B_2-D_2) (B_3^\ast-D_3^\ast)] \nnb \\
&-&\frac{1}{r} m_\ell \,(1 + 3 r - s) 
\,\mbox{\rm Im}[(B_1-D_1) (B_2^\ast-D_2^\ast)] \nnb \\
&+&\frac{1}{r} (1 - r - s)  
\, \mbox{\rm Im}[(B_1^\ast B_4) + (D_1^\ast B_5)] \nnb \\
&-&\frac{1}{r} m_\ell \,(1 - r - s)
\,\mbox{\rm Im}[(B_1-D_1) (B_3^\ast-D_3^\ast)]  \nnb \\
&-&\frac{8}{r} m_B^2 m_\ell \lambda
\,\mbox{\rm Im}[(B_4+B_5)(B_7 C_{TE})^\ast] \\
&+& \frac{16}{r} m_\ell \,(1-r-s) 
\,\mbox{\rm Im}[(B_4+B_5)(B_6 C_{TE})^\ast] \nnb \\
&-& \frac{32}{r} m_\ell \,(1+3 r-s) 
\,\mbox{\rm Im}[(B_4+B_5)(t_1 C_{TE})^\ast] \nnb \\
&-& 16 m_B^2 s \Big(
\,\mbox{\rm Im}[A_1^\ast (C_T-2 C_{TE}) B_6] +
\mbox{\rm Im}[C_1^\ast (C_T+2 C_{TE}) B_6] \Big) \nnb \\
&+& 32 m_B^2 (1-r) \Big(
\,\mbox{\rm Im}[A_1^\ast (C_T-2 C_{TE}) t_1] +
\mbox{\rm Im}[C_1^\ast (C_T+2 C_{TE}) t_1] \Big) \nnb \\
&+& 32 \Big(
\mbox{\rm Im}[B_1^\ast (C_T-2 C_{TE}) t_1]
- \mbox{\rm Im}[D_1^\ast (C_T+2 C_{TE}) t_1] \Big) \nnb \\
&+& 512 m_\ell \,
\ga \vel C_T \ver^2 - 4 \vel C_{TE} \ver^2 \dr
\,\mbox{\rm Im}(B_6^\ast t_1)
\Bigg\}~, \nnb
\eea 
\bea
\label{pnp}
P_N^+&=& \frac{1}{\Delta} \pi v m_B^3 \sqrt{s \lambda} \Bigg\{
- 8 m_\ell \, \mbox{\rm Im}[(B_1^\ast C_1) + (A_1^\ast D_1)] \nnb \\
&+& \frac{1}{r} m_B^2 \lambda 
\,\mbox{\rm Im}[(B_2^\ast B_5) + (D_2^\ast B_4)] \nnb \\
&+& \frac{1}{r} m_B^2 m_\ell \lambda 
\,\mbox{\rm Im}[(B_2-D_2) (B_3^\ast-D_3^\ast)] \nnb \\
&-&\frac{1}{r} m_\ell \,(1 + 3 r - s) 
\,\mbox{\rm Im}[(B_1-D_1) (B_2^\ast-D_2^\ast)] \nnb \\
&-&\frac{1}{r} (1 - r - s)  
\, \mbox{\rm Im}[(B_1^\ast B_5) + (D_1^\ast B_4)] \nnb \\
&-&\frac{1}{r} m_\ell \,(1 - r - s)
\,\mbox{\rm Im}[(B_1-D_1) (B_3^\ast-D_3^\ast)]  \nnb \\
&+&\frac{8}{r} m_B^2 m_\ell \lambda
\,\mbox{\rm Im}[(B_4+B_5)(B_7 C_{TE})^\ast] \\
&-& \frac{16}{r} m_\ell \,(1-r-s) 
\,\mbox{\rm Im}[(B_4+B_5)(B_6 C_{TE})^\ast] \nnb \\
&+& \frac{32}{r} m_\ell \,(1+3 r-s) 
\,\mbox{\rm Im}[(B_4+B_5)(t_1 C_{TE})^\ast] \nnb \\
&-& 16 m_B^2 s \Big(
\,\mbox{\rm Im}[A_1^\ast (C_T+2 C_{TE}) B_6] +
\mbox{\rm Im}[C_1^\ast (C_T-2 C_{TE}) B_6] \Big) \nnb \\
&+& 32 m_B^2 (1-r) \Big(
\,\mbox{\rm Im}[A_1^\ast (C_T+2 C_{TE}) t_1] +
\mbox{\rm Im}[C_1^\ast (C_T-2 C_{TE}) t_1] \Big) \nnb \\
&-& 32 \Big(
\mbox{\rm Im}[B_1^\ast (C_T+2 C_{TE}) t_1]
- \mbox{\rm Im}[D_1^\ast (C_T-2 C_{TE}) t_1] \Big) \nnb \\
&+& 512 m_\ell \,
\ga \vel C_T \ver^2 - 4 \vel C_{TE} \ver^2 \dr
\,\mbox{\rm Im}(B_6^\ast t_1)
\Bigg\}~. \nnb
\eea 

Concerning expressions $P_L^{(\pm)},~P_T^{(\pm)}$ and $P_N^{(\pm)}$ few remarks 
are in order. The difference between $P_L^-$ and $P_L^+$ results from the 
scalar and tensor type interactions. Similar situation takes place for the normal
polarization $P_N^{(\pm)}$ of leptons and antileptons. In the 
$m_\ell \rar 0$ limit, the difference between $P_T^-$ and $P_T^+$ is due to
again existence of new physics, i.e., scalar and tensor type interactions.
For these reasons the experimental study of $P_L^{(\pm)}$ and $P_T^{(\pm)}$
can give essential information about new physics. Note that similar
situation takes place for the inclusive channel $b \rar s \ell^+ \ell^-$
(see \cite{R19}). 

Combined analysis of the lepton and antilepton polarizations can also give
very useful hints in search of new physics, since in the SM $P_L^-+P_L^+=0$,
$P_N^-+P_N^+= 0$ and $P_T^- - P_T^+ \approx 0$.

Using Eqs. (\ref{plm}), (\ref{plp}) we get

\bea
\label{lpl}
P_L^- + P_L^+ &=& \frac{4}{\Delta}  \,m_B^2 v \,\Bigg\{ 
\frac{2}{r} m_\ell \lambda\, 
\mbox{\rm Re} [(B_1 - D_1) (B_4^\ast + B_5^\ast)] \nnb \\
&-& \frac{2}{r} m_B^2 m_\ell \lambda (1-r) \, 
\mbox{\rm Re} [(B_2 - D_2) (B_4^\ast + B_5^\ast)] \nnb \\ 
&-&\frac{1}{r} m_B^2 s \lambda
\Big( \vel B_4 \ver^2 - \vel B_5 \ver^2\Big) -
\frac{2}{r} m_B^2 m_\ell s \lambda \, 
\mbox{\rm Re} [(B_3 - D_3) (B_4^\ast + B_5^\ast)] \nnb \\
&+& \frac{8}{3 r} m_B^4 m_\ell \lambda^2 \,
\mbox{\rm Re} [(B_2 + D_2) (B_7 C_T)^\ast] \nnb \\
&+& \frac{32}{3 r} m_B^6 s  \lambda^2 \vel B_7 \ver^2 
\mbox{\rm Re} (C_T C_{TE}^\ast) \nnb \\
&-& \frac{8}{3 r} m_B^2 m_\ell \lambda (1-r-s) \,
\mbox{\rm Re} [(B_1 + D_1) (B_7 C_T)^\ast] \nnb \\
&-& \frac{16}{3 r} m_B^2 m_\ell \lambda (1-r-s) \,
\mbox{\rm Re} [(B_2 + D_2) (B_6 C_T)^\ast] \nnb \\
&-& \frac{128}{3 r} m_B^4 s \lambda(1-r-s) \,
\mbox{\rm Re} (B_6 B_7^\ast) \, \mbox{\rm Re} (C_T C_{TE}^\ast) \nnb \\
&+& \frac{16}{3 r}  m_\ell (\lambda + 12 r s) \,
\mbox{\rm Re} [(B_1 + D_1) (B_6 C_T)^\ast] \\
&+& \frac{128}{3 r} m_B^2 s (\lambda + 12 r s) \,
\vel B_6 \ver^2 \mbox{\rm Re} (C_T C_{TE}^\ast) \nnb \\
&+& \frac{512}{3 r} m_B^2 \, [ \lambda (4 r + s) + 12 r (1-r)^2 ]
\, \vel t_1 \ver^2 \,\mbox{\rm Re} (C_T C_{TE}^\ast)\nnb \\
&-& \frac{512}{3 r} m_B^2 s \, [\lambda +12 r (1-r) ]\, 
\mbox{\rm Re} (t_1 B_6^\ast) \, \mbox{\rm Re} (C_T C_{TE}^\ast) \nnb \\
&+& \frac{256}{3 r} m_B^4 s \lambda (1+3r -s) \, 
\mbox{\rm Re} (t_1 B_7^\ast) \, \mbox{\rm Re} (C_T C_{TE}^\ast) \nnb \\
&+& \frac{512}{3} m_B^2 m_\ell \lambda \,
\mbox{\rm Re} [(A_1 + C_1) (t_1 C_{TE})^\ast] \nnb \\
&-& \frac{32}{3 r}  m_\ell  \, [\lambda +12 r (1-r) ]\,
\mbox{\rm Re} [(B_1 + D_1) (t_1 C_T)^\ast] \nnb \\
&+& \frac{32}{3 r} m_B^2 m_\ell  \lambda (1+3r -s) \, 
\mbox{\rm Re} [(B_2 + D_2) (t_1 C_T)^\ast]
\Bigg\}~. \nnb
\eea

For the case of transversal polarization, it is the difference of the lepton
and antilepton polarizations that is relevant and it can be calculated  
from Eqs. (\ref{ptm}) and (\ref{ptp}) 
\bea
\label{tmt}
P_T^- - P_T^+ &=& \frac{\pi}{\Delta} m_B \sqrt{s \lambda} \Bigg\{
\frac{2}{r s} m_B^4 m_\ell (1-r) \lambda
\Big[ \vel B_2 \ver^2 - \vel D_2 \ver^2\Big]\nnb \\
&+& \frac{1}{r} m_B^4 \lambda \,
\mbox{\rm Re} [(B_2 + D_2) (B_4^\ast - B_5^\ast)] \nnb \\
&+& \frac{2}{r} m_B^4 m_\ell \lambda \,                  
\mbox{\rm Re} [(B_2 + D_2) (B_3^\ast - D_3^\ast)] \nnb \\
&+& \frac{2}{r} m_B^2 m_\ell (1+3 r - s)  \, 
\Big[ \mbox{\rm Re}(B_1 D_2^\ast) -  \mbox{\rm Re}(B_2 D_1^\ast)\Big] \nnb \\
&+& \frac{2}{rs} m_\ell (1-r-s)                    
\Big[ \vel B_1 \ver^2 - \vel D_1 \ver^2\Big]\nnb \\
&-& \frac{1}{r} m_B^2 (1-r-s)                   
\mbox{\rm Re} [(B_1 + D_1) (B_4^\ast - B_5^\ast)] \nnb \\
&-& \frac{2}{r} m_B^2 m_\ell (1-r-s)            
\mbox{\rm Re} [(B_1 + D_1) (B_3^\ast - D_3^\ast)] \nnb \\
&-&\frac{2}{r s} m_B^2 m_\ell [\lambda + (1-r) (1-r-s)]
\Big[\mbox{\rm Re}(B_1 B_2^\ast) - \mbox{\rm Re}(D_1 D_2^\ast)\Big] \nnb \\
&-&\frac{32}{r s} m_B^2 m_\ell^2 \lambda \,
\mbox{\rm Re}[(B_1-D_1) (B_7 C_{TE})^\ast] \nnb \\
&+&\frac{32}{r s} m_B^4 m_\ell^2 \lambda (1-r)\,
\mbox{\rm Re}[(B_2-D_2) (B_7 C_{TE})^\ast] \\
&+&\frac{16}{r} m_B^4 m_\ell \lambda \,  
\mbox{\rm Re}[(B_4-B_5) (B_7 C_{TE})^\ast] \nnb \\
&+&\frac{32}{r} m_B^4 m_\ell^2 \lambda \,       
\mbox{\rm Re}[(B_3-D_3) (B_7 C_{TE})^\ast] \nnb \\
&+&\frac{64}{r s} m_\ell^2 (1-r-s) \,           
\mbox{\rm Re}[(B_1-D_1) (B_6 C_{TE})^\ast] \nnb \\
&-&\frac{64}{r s} m_B^2 m_\ell^2 (1-r)(1-r-s) \,           
\mbox{\rm Re}[(B_2-D_2) (B_6 C_{TE})^\ast] \nnb \\
&-&\frac{32}{r} m_B^2 m_\ell (1-r-s) \,           
\mbox{\rm Re}[(B_4-B_5) (B_6 C_{TE})^\ast] \nnb \\
&-&\frac{64}{r} m_B^2 m_\ell^2 (1-r-s) \,           
\mbox{\rm Re}[(B_3-D_3) (B_6 C_{TE})^\ast] \nnb \\
&+& 32 m_B^4 s v^2 \,           
\mbox{\rm Re}[(A_1-C_1) (B_6 C_T)^\ast] \nnb \\
&+&\frac{64}{r} m_B^2 m_\ell (1+3 r-s) \,           
\mbox{\rm Re}[(B_4-B_5) (t_1 C_{TE})^\ast] \nnb \\
&-& 64 m_B^4 (1-r) v^2 \,                      
\mbox{\rm Re}[(A_1-C_1) (t_1 C_T)^\ast] \nnb \\
&+&\frac{128}{r s} [m_B^2 r s - m_\ell^2 (1+7 r-s)] \,
\mbox{\rm Re}[(B_1-D_1) (t_1 C_{TE})^\ast] \nnb \\
&+&\frac{128}{r s} m_B^2 m_\ell^2 (1-r) (1+3 r-s)
\mbox{\rm Re}[(B_2-D_2) (t_1 C_{TE})^\ast] \nnb \\
&+&\frac{128}{r} m_B^2 m_\ell^2 (1+3 r-s)
\mbox{\rm Re}[(B_3-D_3) (t_1 C_{TE})^\ast]
\Bigg\}~. \nnb
\eea

In the same manner it follows from Eqs. (\ref{pnm}) and (\ref{pnp})
\bea
\label{npn}
P_N^- + P_N^+ &=& \frac{1}{\Delta} \pi v m_B^3 \sqrt{s\lambda} \Bigg\{
- \frac{2}{r} m_\ell (1+3 r -s) \,
\mbox{\rm Im} [(B_1 - D_1) (B_2^\ast - D_2^\ast)] \nnb \\
&-& \frac{2}{r} m_\ell (1-r -s) \,
\mbox{\rm Im} [(B_1 - D_1) (B_3^\ast - D_3^\ast)] \nnb \\
&-& \frac{1}{r} (1-r -s) \,
\mbox{\rm Im} [(B_1 - D_1) (B_4^\ast - B_5^\ast)] \nnb \\
&+&\frac{2}{r} m_B^2  m_\ell \lambda \,
\mbox{\rm Im} [(B_2 - D_2) (B_3^\ast - D_3^\ast)] \\
&+&\frac{1}{r} m_B^2 \lambda \, 
\mbox{\rm Im} [(B_2 - D_2) (B_4^\ast - B_5^\ast)] \nnb \\
&+& 32 m_B^2 s \, \mbox{\rm Im} [(A_1 + C_1)(B_6 C_T)^\ast] \nnb \\
&+&1024 m_\ell \Big(  \vel C_T \ver^2  - \vel 4 C_{TE} \ver^2  \Big)
\mbox{\rm Im} (B_6^\ast t_1) \nnb \\
&-& 64 m_B^2 (1-r) \, \mbox{\rm Im} [(A_1 + C_1)(t_1 C_T)^\ast] \nnb \\
&+& 128 \, \mbox{\rm Im} [(B_1 + D_1)(t_1 C_{TE})^\ast]
\Bigg\}~. \nnb 
\eea
It is evident from Eq. (\ref{lpl}) that the "pure" SM contribution to the 
$P_L^-+P_L^+$ completely disappears. Therefore a measurement of the nonzero
value of $P_L^-+P_L^+$ in future experiments, is an indication of the
discovery of new physics beyond SM.

\section{Numerical analysis}
The input parameters we used in our analysis are:
$\vel V_{tb} V_{ts}^\ast \ver = 0.0385$, $\alpha^{-1}=129$,
$G_F=1.17\times 10^{-5}~GeV^{-2}$, $~\Gamma_B=4.22\times10^{-13}~GeV$,
$C_9^{eff}=4.344,~C_{10}=-4.669$. 
It should be noted here that the above--value of the Wilson coefficient
$C_9^{eff}$
we have used in our numerical calculations corresponds only to short
distance contribution. In addition to the short distance contribution
$C_9^{eff}$
also receives long distance contributions associated with the real 
$\bar c c$ intermediate states, i.e., with the $J/\psi$ family. In this work
we restricted ourselves only to short distance contributions. As far as
$C_7^{eff}$ is concerned, experimental results fixes only the modulo of it.
For this reason throughout our analysis we have considered both
possibilities, i.e., $C_7^{eff} = \mp 0.313$, where the upper sign
corresponds to the SM prediction. The values of the input parameters which
are summarized above, have been fixed by their central values.  
 
For the values of the form factors, we have used the results of
\cite{R28}, where  the radiative corrections to the leading twist
contribution and $SU(3)$ breaking effects are also taken into account.
The $q^2$ dependence of the form factors can be represented in terms of
three parameters as
\bea
F(q^2) = \frac{F(0)}{\ds 1-a_F\,\frac{q^2}{m_B^2} + b_F \left
    ( \frac{q^2}{m_B^2} \right)^2}~, \nnb
\eea
where the values of parameters $F(0)$, $a_F$ and $b_F$ for the
$B \rar K^\ast$ decay are listed in Table 1. Note that in the present analysis
the final state Coulomb interactions of the leptons with the other charged
particles are neglected since this effect is known to be much smaller than
the averaged values of the SM (see \cite{R24}). Furthermore the final state
interaction of the lepton polarization for the 
$K_L \rar \pi^+ \mu^- \bar \nu_\mu$ or $K^+ \rar \pi^+ \mu^- \mu^+$ decays is
estimated to be of the order of $\alpha(m_\mu/m_K) \approx 10^{-3}$
\cite{R29}. For this reason the final state interaction effect is neglected
as well.  

\begin{table}[h]                    
\renewcommand{\arraystretch}{1.5}                        
\addtolength{\arraycolsep}{3pt}
$$
\begin{array}{|l|ccc|}
\hline
& F(0) & a_F & b_F \\ \hline
A_1^{B \rar K^*} &
\phantom{-}0.34 \pm 0.05 & 0.60 & -0.023 \\
A_2^{B \rar K^*} &
\phantom{-}0.28 \pm 0.04 & 1.18 & \phantom{-}0.281\\
V^{B \rar K^*} &
 \phantom{-}0.46 \pm 0.07 & 1.55 & \phantom{-}0.575\\
T_1^{B \rar K^*} &
  \phantom{-}0.19 \pm 0.03 & 1.59 & \phantom{-}0.615\\
T_2^{B \rar K^*} & 
 \phantom{-}0.19 \pm 0.03 & 0.49 & -0.241\\
T_3^{B \rar K^*} & 
 \phantom{-}0.13 \pm 0.02 & 1.20 & \phantom{-}0.098\\ \hline
\end{array}   
$$
\caption{$B$ meson decay form factors in a three-parameter fit, where the
radiative corrections to the leading twist contribution and SU(3) breaking
effects are taken into account.}
\renewcommand{\arraystretch}{1}
\addtolength{\arraycolsep}{-3pt}
\end{table}       
We observe from the explicit form of the expressions of the lepton
polarizations that they all depend on $q^2$ and the new Wilson coefficients.
Therefore it may be experimentally difficult to study the dependence of the
the polarizations of each lepton on all $\ell^+ \ell^-$ center of mass
energies and on new Wilson coefficients. So we eliminate the dependence of
the lepton polarizations on one of the variables, namely $q^2$, by
performing integration over $q^2$ in the allowed kinematical region, so that
the lepton polarizations are averaged. The averaged lepton polarizations are
defined as
\bea
\label{av}
\lla P_i \rra = \frac{\ds \int_{4 m_\ell^2}^{(m_b-m_{K^\ast})^2}
P_i \frac{d{\cal B}}{dq^2} dq^2}
{\ds \int_{4 m_\ell^2}^{(m_b-m_{K^\ast})^2}
 \frac{d{\cal B}}{dq^2} dq^2}~.  
\eea
 
We present our analysis in a series Figures. Figs. (1) and (2) depict the
dependence of the averaged longitudinal polarization $\lla P_L^- \rra$ of
$\ell^-$ and the combination $\lla P_L^- + P_L^+ \rra$ on new Wilson coefficients, 
at $C_7^{eff}=-0.313$ for $B \rar K^\ast \mu^+ \mu^-$ decay. From these figures we
observe that $\lla P_L^- \rra$ is more sensitive to the existence of the
tensor interaction, while the combined average $\lla P_L^- + P_L^+ \rra$ is
to both scalar and tensor type interactions. As has already been noted, 
this is an expected result since vector type interactions are canceled when
the combined longitudinal polarization asymmetry of the lepton and
antilepton is considered.
From Fig. (2) we see that 
$\lla P_L^- + P_L^+ \rra=0$ at $C_X=0$, which confirms the SM result as
expected.
For the other choice of $C_7^{eff}$, i.e., 
$C_7^{eff}=0.313$ the situation 
is similar to the previous case, but the magnitude of $\lla P_L^- + P_L^+ \rra$ 
is smaller. Figs. (3) and (4) are the same as Figs.(1) and (2) but for the
$B \rar K^\ast \tau^+ \tau^-$ decay. Similar to the muon longitudinal
polarization, $\lla P_L^- \rra$ is strongly dependent on the tensor
interaction coefficients $C_T$ and $C_{TE}$. It is very interesting to
observe that for $C_{TE}>0.5$ $\lla P_L^- \rra$ changes sign, but for all
other cases $\lla P_L^- \rra$ is negative.         

From Fig. (4) we see that the dependence of $\lla P_L^- + P_L^+ \rra$ on
$C_T$ is stronger. Furthermore if the values of the new Wilson coefficients
$C_{LRRL},~C_{LRLR}$ and $C_T$ are negative (positive) so is $\lla P_L^-
\rra$ negative (positive). The situation is to the contrary for the
coefficients $C_{RLRL},~C_{RLLR}$, i.e., $\lla P_L^- + P_L^+ \rra$ is
positive (negative) when the corresponding Wilson coefficients are negative
(positive). Absolutely similar situation takes place for $C_7^{eff}>0$. 
For these reasons determination of the sign and of course magnitude of 
$\lla P_L^- \rra$ and $\lla P_L^- + P_L^+\rra$ can give promising
information about new physics. 

In Figs. (5) and (6) the dependence of the average transversal polarization
$\lla P_T^- \rra$ and the combination $\lla P_T^- - P_T^+ \rra$ on the new
Wilson coefficients, respectively, for the $B \rar K^\ast \mu^+ \mu^-$
decay and at $C_7^{eff}=-0.313$. We observe from Fig. (5) that the average
transversal polarization is strongly dependent on $C_T,~C_{TE},~C_{LRRL}$
and $C_{RLRL}$ and quite weakly to remaining Wilson coefficients. It is also
interesting to note that for the negative (positive) values of the
coefficients $C_{TE}$ and $C_{LRRL}$, $\lla P_T^- \rra$ is negative (positive)
while it follows the opposite path for the coefficients $C_T$ and
$C_{RLRL}$. For the $\lla P_T^- - P_T^+ \rra$ case, there appears strong
dependence on the tensor interactions $C_T$ and $C_{TE}$, as well as all
four scalar interactions with coefficients
$C_{LRRL},~C_{RLLR},~C_{LRLR},~C_{RLRL}$. The behavior of this combined
average is identical for the coefficients $C_{LRLR},~C_{RLRL}$ and 
$C_{LRRL},~C_{RLLR}$ in pairs, so that four lines responsible for these interactions
appear only to be two. Moreover $\lla P_T^- - P_T^+ \rra$ is negative
(positive) for the negative (positive) values of the new Wilson coefficients 
$C_{TE},~C_{LRRL}$ and $C_{RLLR}$. The situation is the other way around for
the coefficients $C_T,~C_{LRLR}$ and $C_{RLRL}$.   
Remembering that in SM, in massless lepton case $\lla P_T^- \rra \approx 0$ and 
$\lla P_T^- - P_T^+ \rra \approx 0$, determination of the signs of the 
$\lla P_T^- \rra$ and $\lla P_T^- - P_T^+ \rra$ can give quite a useful
information about the existence of new physics. For the choice of
$C_7^{eff}=0.313$, apart from the minor differences in their magnitudes, the behaviors 
of $\lla P_T^- \rra$ and $\lla P_T^- - P_T^+ \rra$
are similar as in the previous case. 

As is obvious from Figs. (7) and (8), $\lla P_T^- \rra$ shows stronger dependence on $C_T$
and $\lla P_T^- - P_T^+ \rra$ on $C_T$ and $C_{TE}$, respectively, at
$C_7^{eff}=-0.313$ for the $B \rar K^\ast \tau^+ \tau^-$ decay. 
Again change in signs of $\lla P_T^- \rra$ and  $\lla P_T^- - P_T^+ \rra$ 
are observed depending on the change in the tensor interaction
coefficients. As has already been noted, determination of the sign and
magnitude of  $\lla P_T^- \rra$ and $\lla P_T^- - P_T^+ \rra$     
are useful tools in looking for new physics.

Note that for simplicity all new Wilson coefficients in this work are 
assumed to be real. Under this condition $\lla P_N^- \rra$ and $\lla P_N^- + P_N^+ \rra$
have non-vanishing values coming from the imaginary part of SM, i.e., from
$C_9^{eff}$. From Fig. (9) we see that  $\lla P_N^- \rra$ is strongly
dependent on all tensor and scalar type interactions. On the other hand 
Fig. (10) depicts that the behavior $\lla P_N^- + P_N^+ \rra$ is determined
by only the tensor interactions, for $B \rar K^\ast \mu^+ \mu^-$ decay. 
Similar behavior takes place for the $B \rar K^\ast \tau^+ \tau^-$ decay as
well, as can easily be seen in Figs. (11) and (12). The change in sign and 
magnitude of both $\lla P_N^- \rra$ and $\lla P_N^- + P_N^+ \rra$ that are 
observed in these figures is an indication of the fact that an experimental 
verification of them can give unambiguous information about new physics.

In Figs. (13), (14) and (15) we present parametric plot of the correlations between 
the integrated branching ratio and averaged lepton polarization asymmetries of 
$\tau^-$ and $\tau^+$ as a function of the new Wilson coefficients. 
In Fig. (13) we present the flows in the 
$\ga {\cal B},~\lla P_L^- + P_L^+ \rra \dr$ plane by varying the coefficients of
the tensor and scalar type interactions. Fig. (14) shows the flows in 
$\ga {\cal B},~\lla P_T^- - P_T^+ \rra \dr$ plane by varying the coefficients of
vector, scalar and tensor type interactions. Finally, Fig. (15) depicts the
flows in $\ga {\cal B},\lla P_N^- + P_N^+ \rra \dr$ plane by changing the
coefficients of the tensor type interactions only. 

It should be noted that the influence of the variation of various
coefficients confirms our previous results, i.e., the influence of the
tensor interactions is quite large. The ranges of variation of the new Wilson
coefficients are determined by assuming that the value of the branching
ratio is about the SM prediction. For example if branching ratio is
restricted to have the values in the range 
$10^{-7} \le {\cal B} (B \rar K^\ast \tau^+ \tau^-) \le 5 \times 10^{-7}$,
then it follows from Fig. (13) that the new Wilson coefficients of the
tensor interactions lie in the region $-2.6 \le C_T \le 1.55$ or 
$-0.35 \le C_{TE} \le 1.15$, while all scalar interaction coefficients vary 
in the range between $-4$ and $4$ (in the present work we vary all
coefficients in the range $-4$ and $4$).

Finally we would like to discuss briefly the detectibilty of the lepton
polarization asymmetries. Experimentally, to be able to measure an
asymmetry $\lla P_i \rra$ of a decay with the branching ratio $B$ at the 
$n \sigma$, the required number of events are $N = n^2 / ({\cal B} \lla P_i
\rra^2$. As an example for detecting $\lla P_T \rra \simeq 0.3$ the number
of events expected is $N \simeq 6 \times 10^7 n^2$ events. Therefore at
$B$ factories detection of polarization asymmetries for $\tau$ could be
accessible. 

\section{Summary and Conclusions}

In this work we present the most general analysis of the lepton polarization
asymmetries in the rare $B \rar K^\ast \ell^+ \ell^-~(\ell=\mu,~\tau)$  
using the general, model independent form of the effective Hamiltonian.
The dependence of the longitudinal, transversal and normal polarization
asymmetries of $\ell^+$ and $\ell^-$ and their combined asymmetries 
on the new Wilson coefficients are
studied. It is found that the lepton polarization asymmetries are very
sensitive to the existence tensor and scalar type interactions. Moreover,
$\lla P_T \rra$ and $\lla P_N \rra$ change their signs 
for the $B \rar K^\ast \mu^+ \mu^-$ and $\lla P_L \rra$ and $\lla P_T \rra$
change their signs for the $B \rar K^\ast \tau^+ \tau^-$ decays,
respectively, as the new
Wilson coefficients vary in the region of interest.
This conclusion is valid also for the
combined polarization effects 
$\lla P_L^- + P_L^+ \rra,~\lla P_T^- - P_T^+ \rra$ and 
$\lla P_N^- + P_N^+ \rra$ for the same decay channel. It is well known that
in the SM, $\lla P_L^- + P_L^+ \rra = 
\lla P_T^- - P_T^+ \rra = \lla P_N^- + P_N^+ \rra \simeq 0$           
in the limit $m_\ell \rar 0$. Therefore any deviation from this relation and
determination of the sign of polarization is decisive and effective tool in
looking for new physics beyond SM.

\newpage
\section*{Figure captions}
{\bf Fig. (1)} The dependence of the average longitudinal polarization asymmetry
$\lla P_L^- \rra$ of $\ell^-$ on the new Wilson coefficients 
at $C_7^{eff}=-0.313$ for the $B \rar K^\ast \mu^- \mu^+$ decay. \\ \\
{\bf Fig. (2)} The dependence of the combined average longitudinal
polarization asymmetry $\lla P_L^- + P_L^+\rra$ of $\ell^-$ and $\ell^+$ 
on the new Wilson coefficients
at $C_7^{eff}=-0.313$ for the $B \rar K^\ast \mu^- \mu^+$ decay. \\ \\ 
{\bf Fig. (3)} The same as in Fig. (1), but for the 
$B \rar K^\ast \tau^- \tau^+$ decay. \\ \\
{\bf Fig. (4)} The same as in Fig. (2), but for the 
$B \rar K^\ast \tau^- \tau^+$ decay. \\ \\
{\bf Fig. (5)} The same as in Fig. (1), but for the average transversal
polarization asymmetry $\lla P_T^- \rra$ of $\ell^-$. \\ \\
{\bf Fig. (6)} The same as in Fig. (2), but for the transversal polarization
asymmetry $\lla P_T^- - P_T^+\rra$. \\ \\
{\bf Fig. (7)} The same as in Fig. (5), but for the 
$B \rar K^\ast \tau^- \tau^+$ decay. \\ \\
{\bf Fig. (8)} The same as in Fig. (6), but for the 
$B \rar K^\ast \tau^- \tau^+$ decay. \\ \\
{\bf Fig. (9)} The dependence of the average normal asymmetry
$\lla P_N^- \rra$ of $\ell^-$ on the new Wilson coefficients
at $C_7^{eff}=-0.313$ for the $B \rar K^\ast \mu^- \mu^+$ decay. \\ \\
{\bf Fig. (10)} The dependence of the combined average normal polarization 
asymmetry $\lla P_N^- + P_N^+\rra$ of $\ell^-$ and $\ell^+$
on the new Wilson coefficients
at $C_7^{eff}=-0.313$ for the $B \rar K^\ast \mu^- \mu^+$ decay. \\ \\
{\bf Fig. (11)} The same as in Fig. (9), but for the 
$B \rar K^\ast \tau^- \tau^+$ decay. \\ \\
{\bf Fig. (12)} The same as in Fig. (10), but for the 
$B \rar K^\ast \tau^- \tau^+$ decay. \\ \\
{\bf Fig. (13)} Parametric plot of the correlation between the integrated
branching ratio ${\cal B}$ (in units of $10^{-7}$) and the combined  
average longitudinal lepton
polarization asymmetry $\lla P_L^- + P_L^+ \rra$
at $C_7^{eff}=-0.313$ as function of the new Wilson coefficients as indicated
in the figure, for the $B \rar K^\ast \tau^- \tau^+$ decay. \\ \\ 
{\bf Fig. (14)} The same as in Fig. (13), but for the 
combined average transversal lepton polarization asymmetry $\lla P_T^- - P_T^+\rra$ . \\ \\
{\bf Fig. (15)} The same as in Fig. (13), but for the combined average normal
lepton polarization asymmetry $\lla P_N^- + P_N^+\rra$.

\newpage

\begin{figure}
\vskip 1cm
    \includegraphics{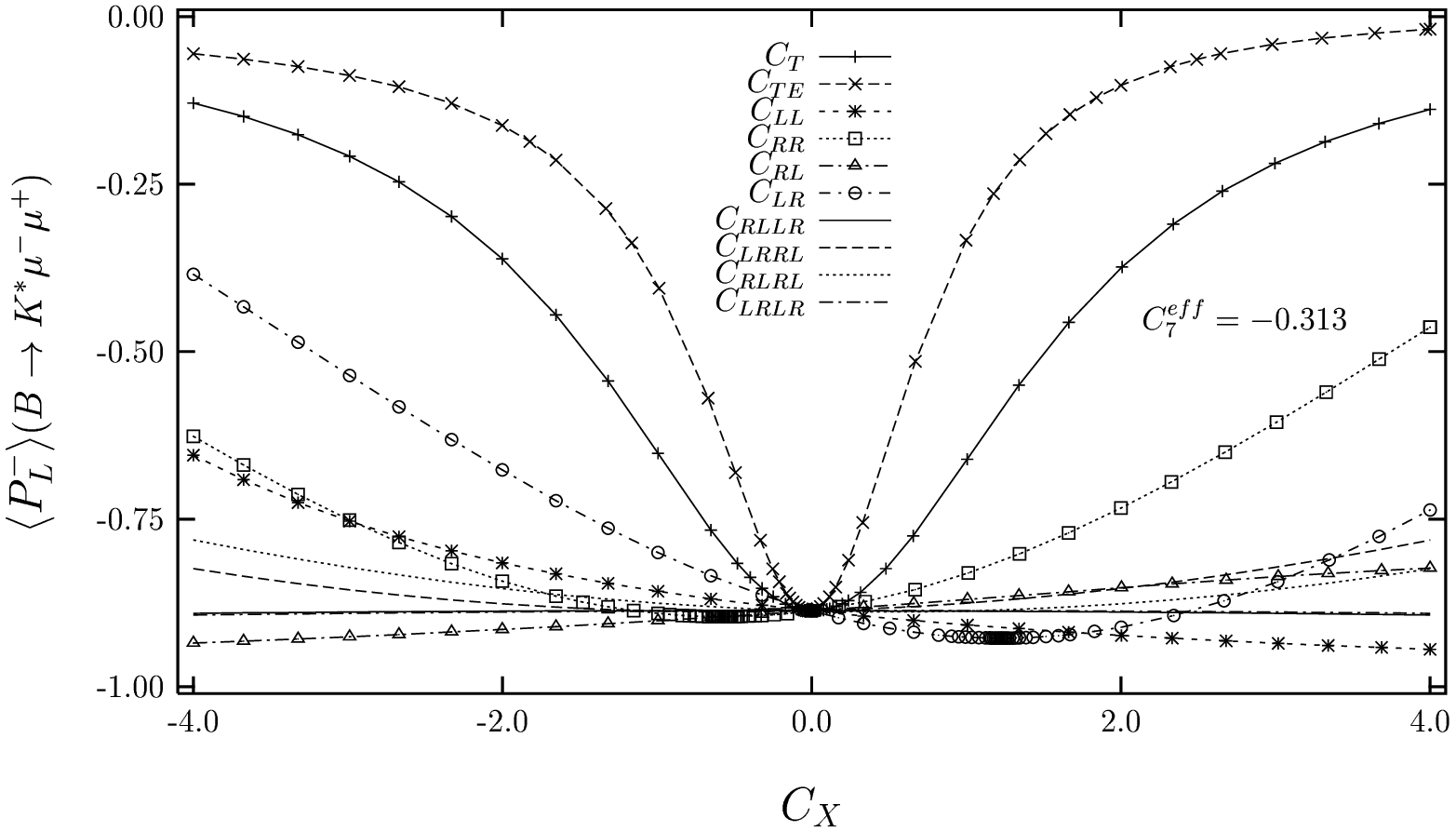}
\vskip 8.1cm
\caption{}
\end{figure}

\begin{figure}
\vskip 1.5 cm
    \includegraphics{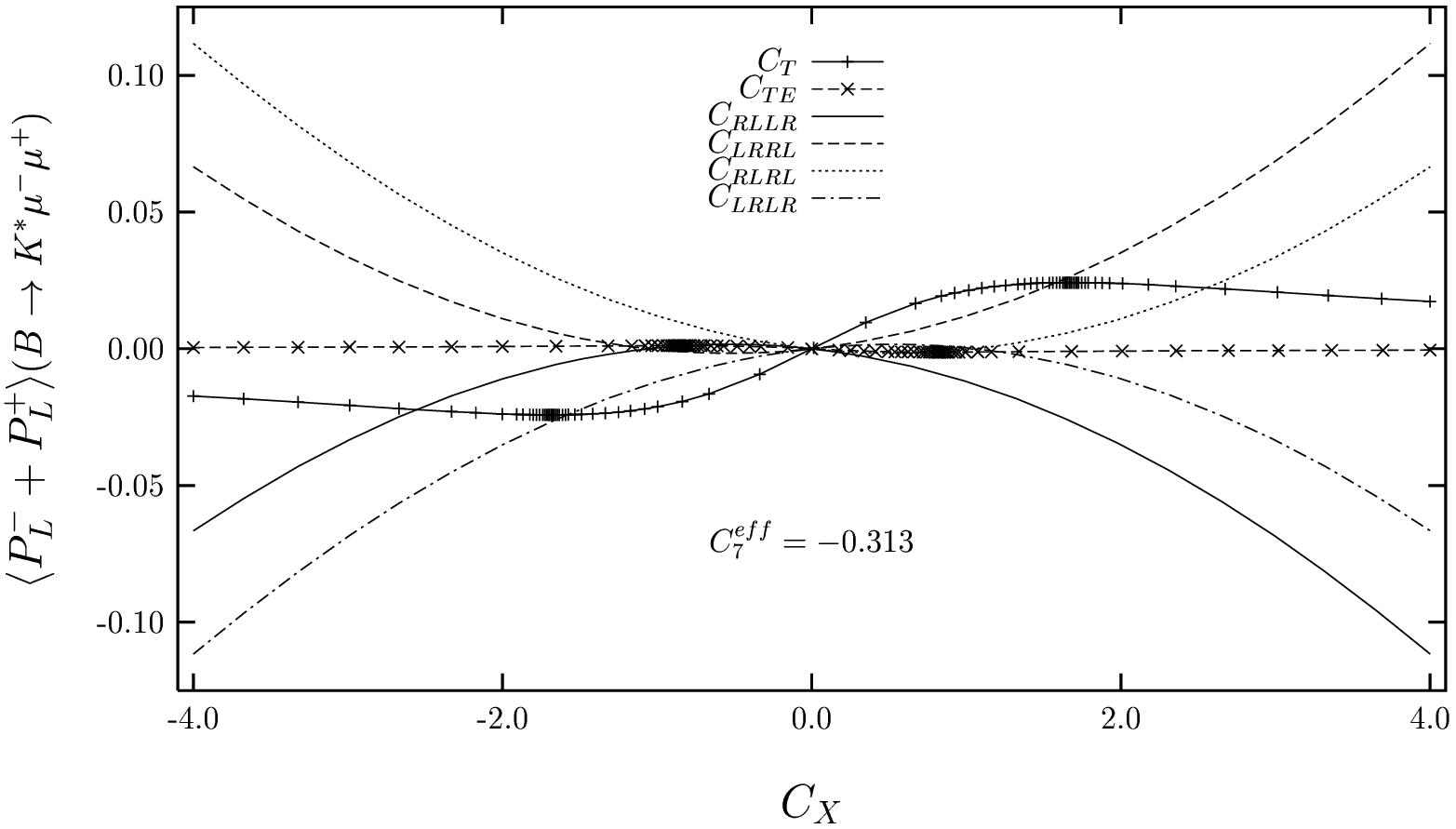}
\vskip 9. cm
\caption{}
\end{figure}

\begin{figure}
\vskip 1.5 cm
    \includegraphics{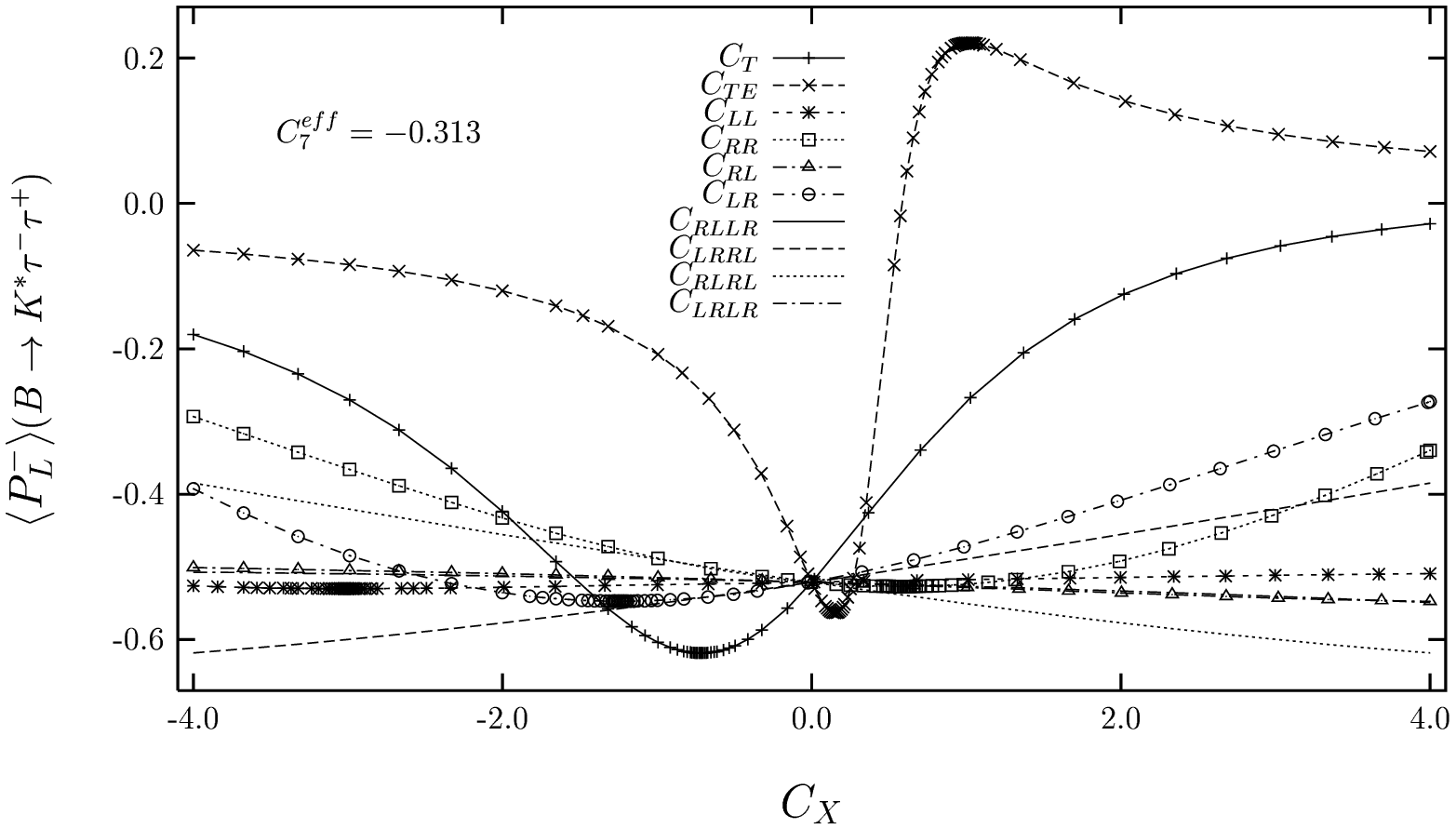}
\vskip 9. cm
\caption{}
\end{figure}

\begin{figure}
\vskip 1cm
    \includegraphics{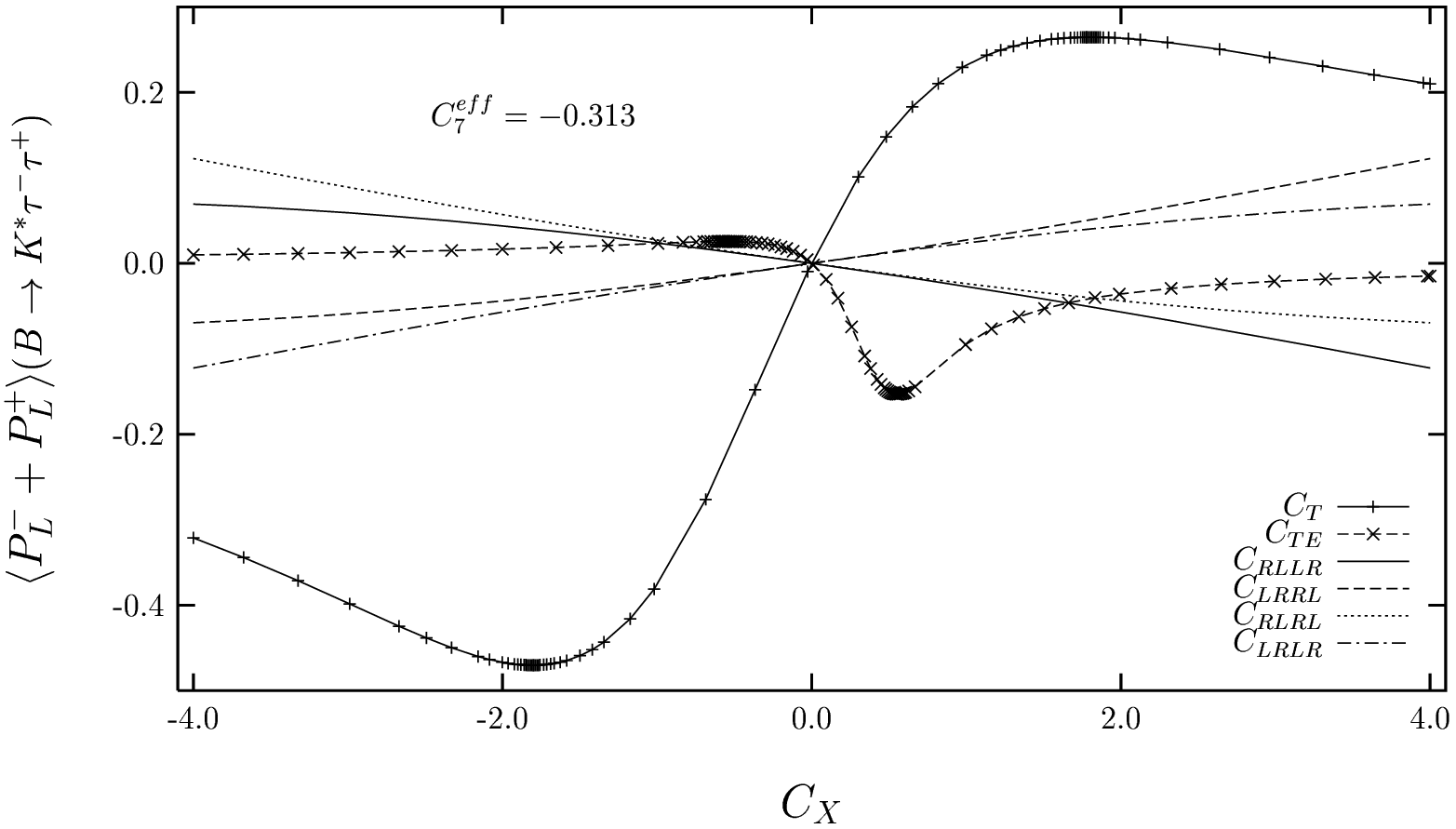}
\vskip 8.1cm
\caption{}
\end{figure}

\begin{figure}
\vskip 1.5 cm
    \includegraphics{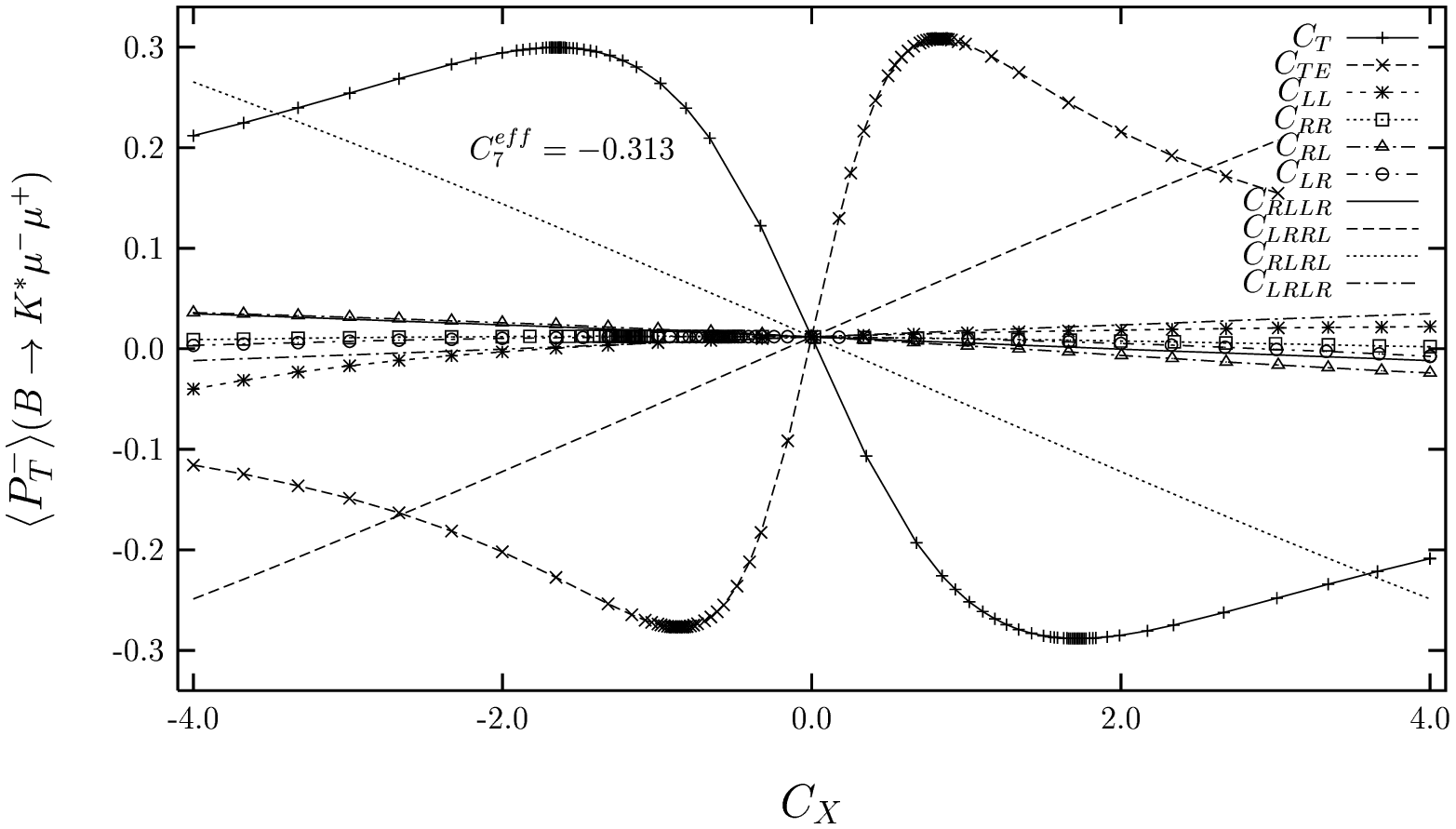}
\vskip 9. cm
\caption{}
\end{figure}

\begin{figure}
\vskip 1cm
    \includegraphics{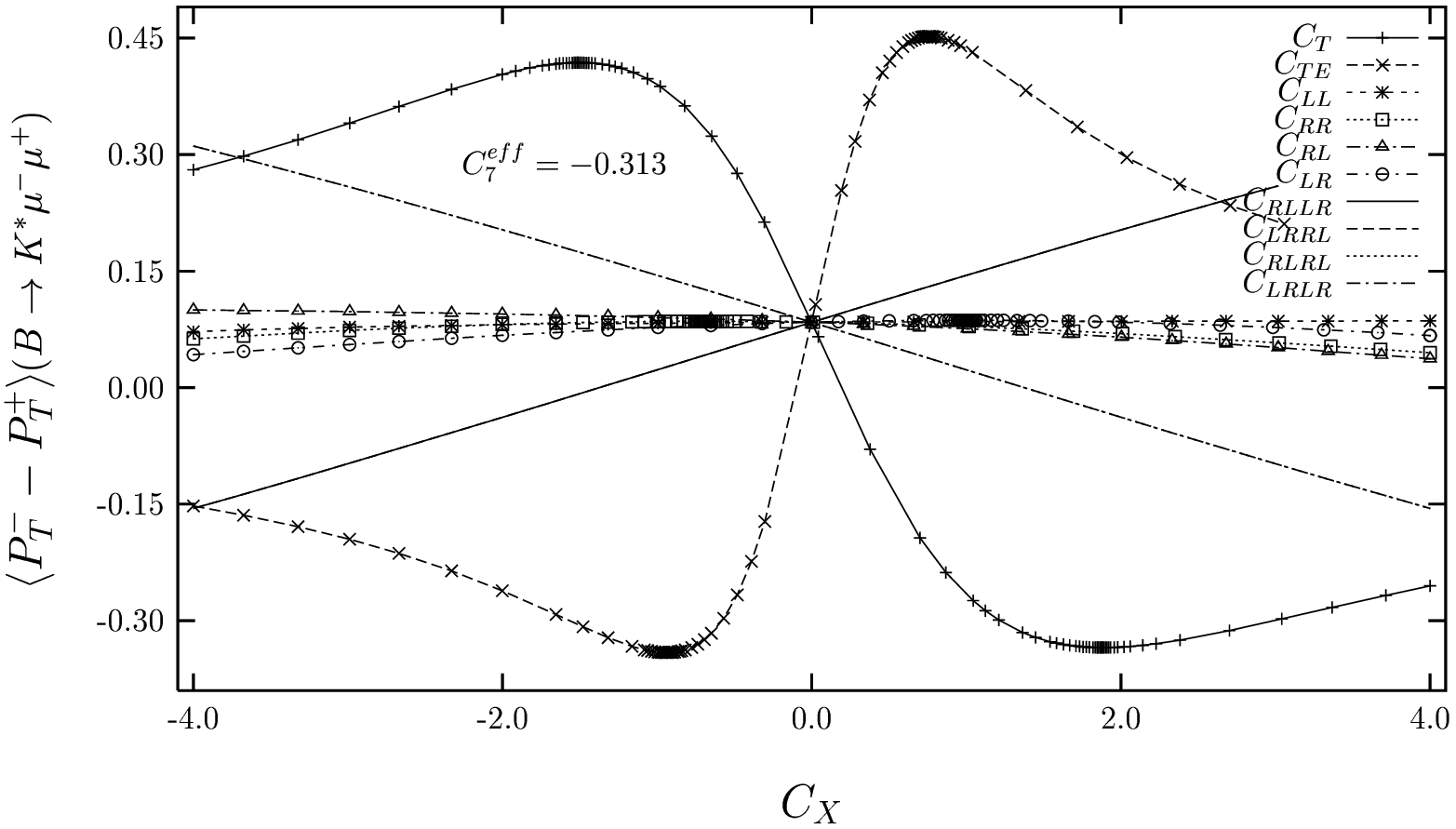}
\vskip 8.1cm
\caption{}
\end{figure}

\begin{figure}
\vskip 1.5 cm
    \includegraphics{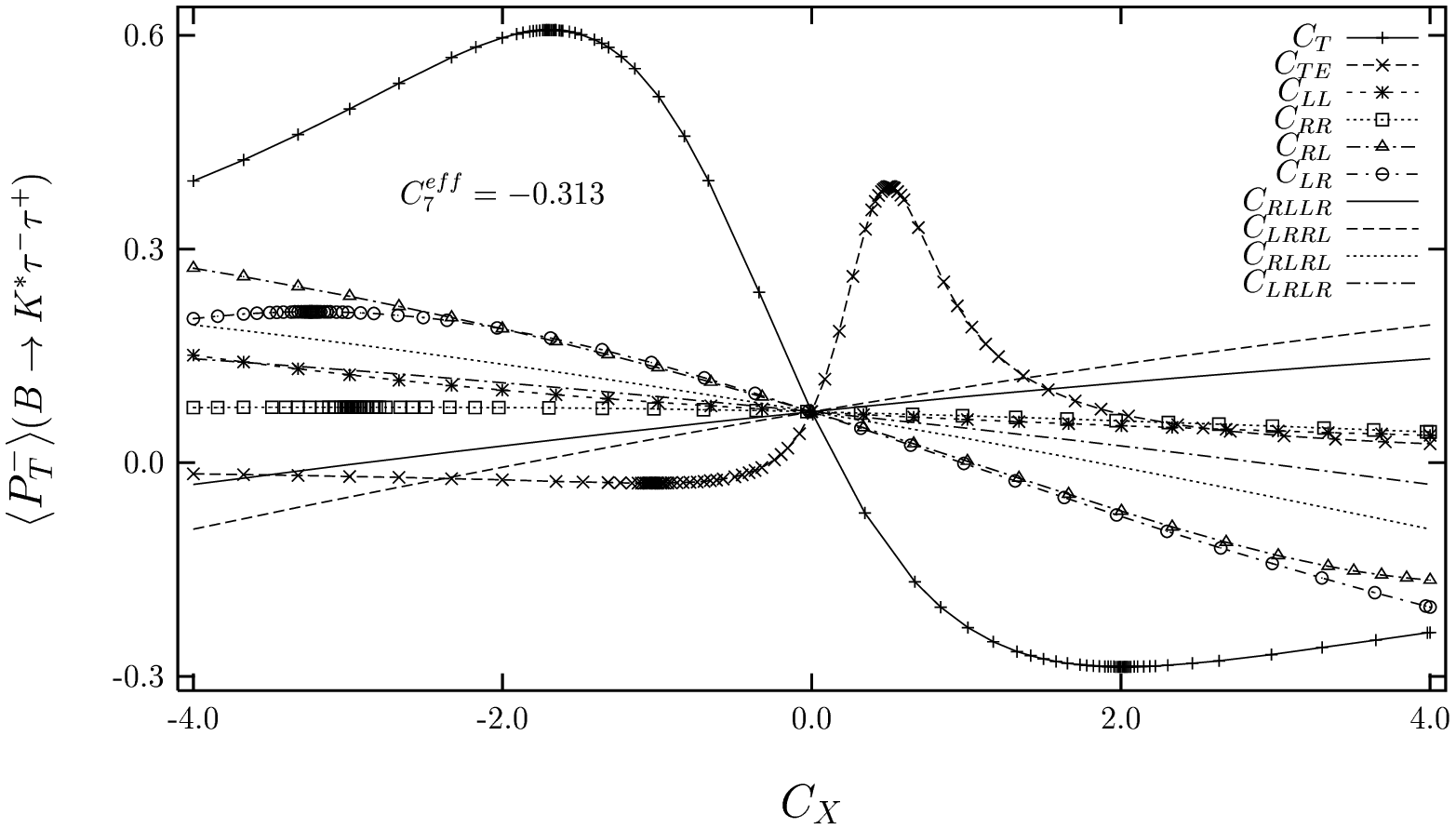}
\vskip 9. cm
\caption{}
\end{figure}

\begin{figure}
\vskip 1cm
    \includegraphics{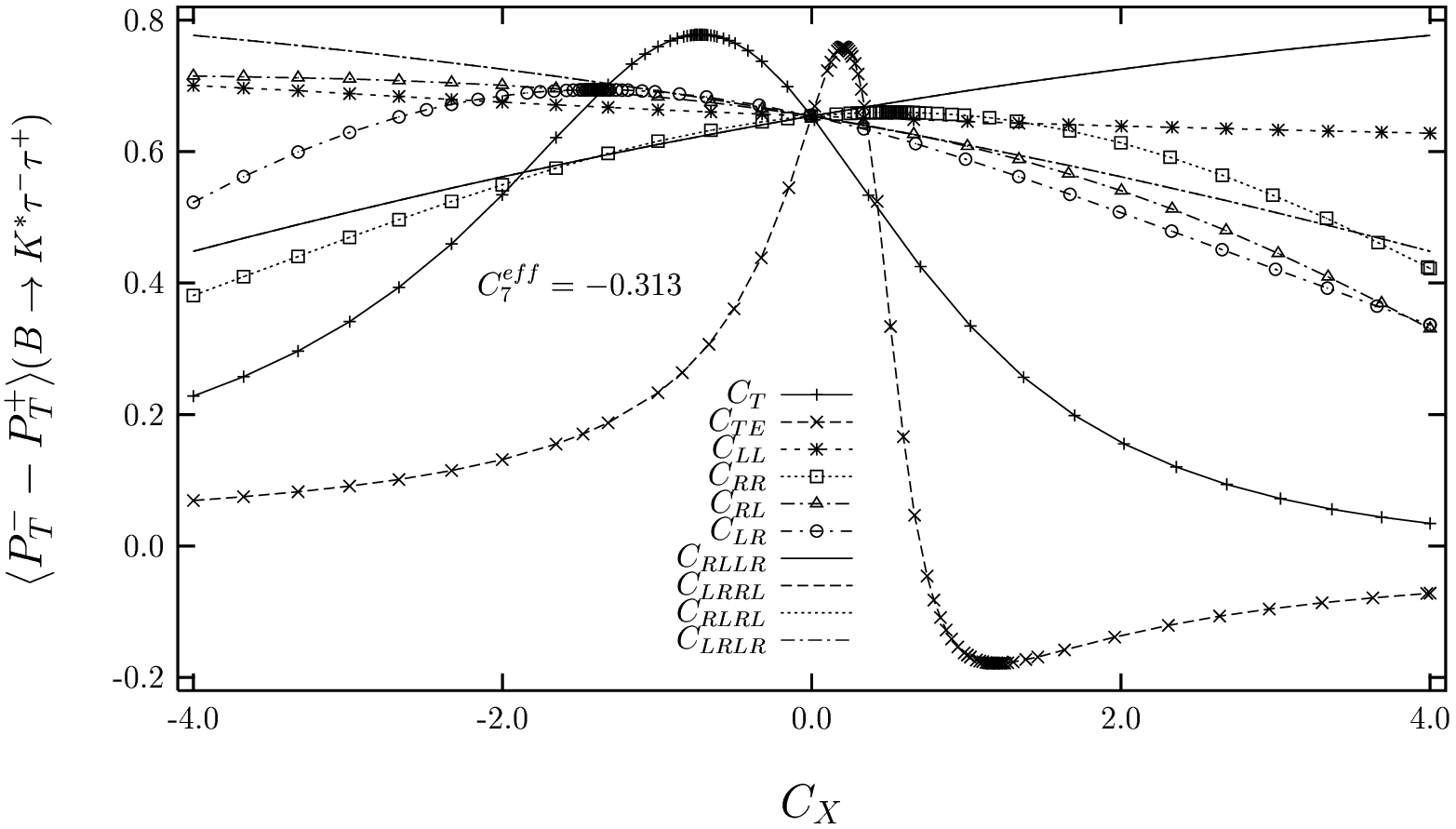}
\vskip 8.1cm
\caption{}
\end{figure}

\begin{figure}
\vskip 1.5 cm
    \includegraphics{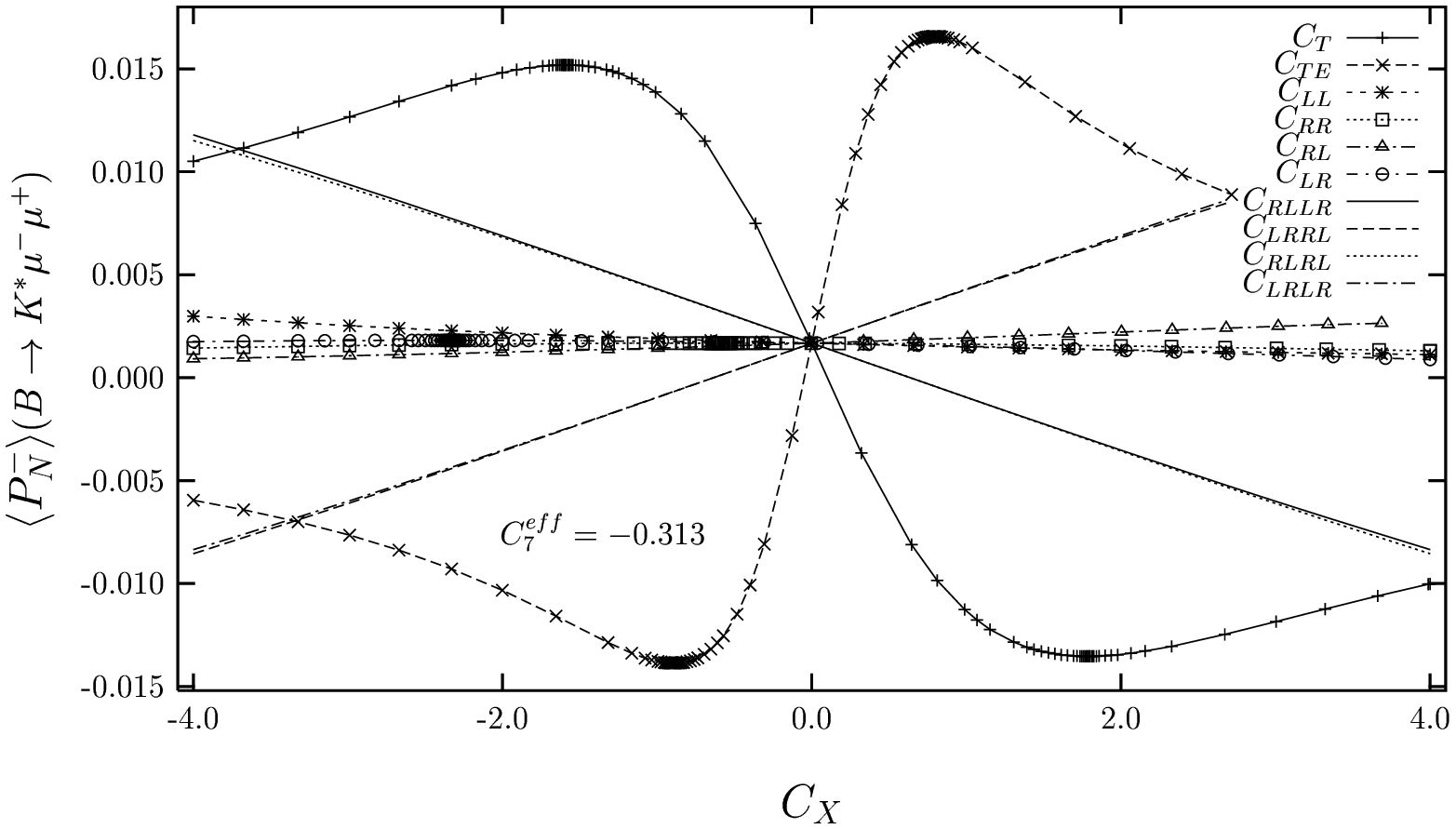}
\vskip 9. cm
\caption{}
\end{figure}

\begin{figure}
\vskip 1cm
    \includegraphics{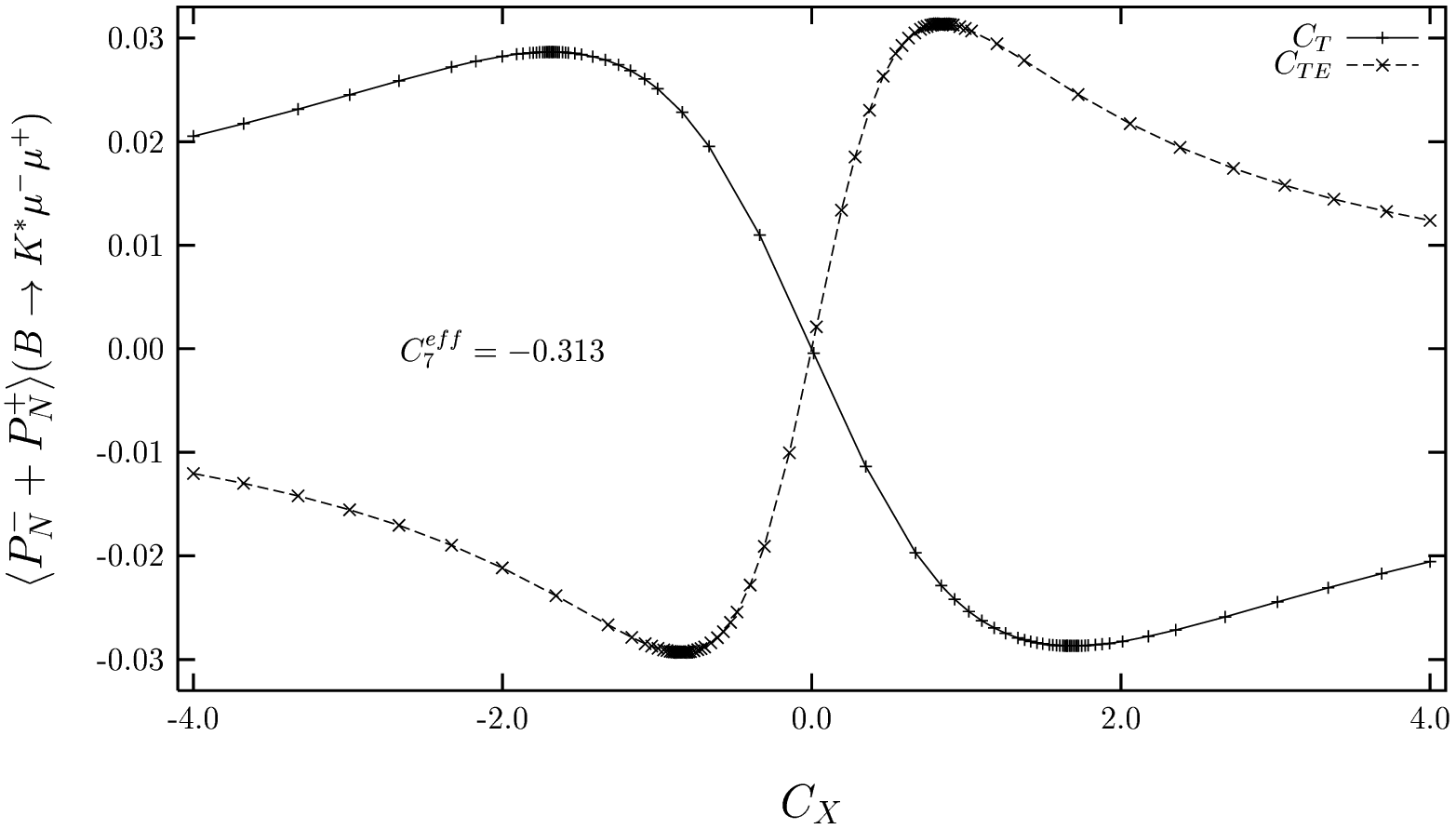}
\vskip 8.1cm
\caption{}
\end{figure}

\begin{figure}
\vskip 1.5 cm
    \includegraphics{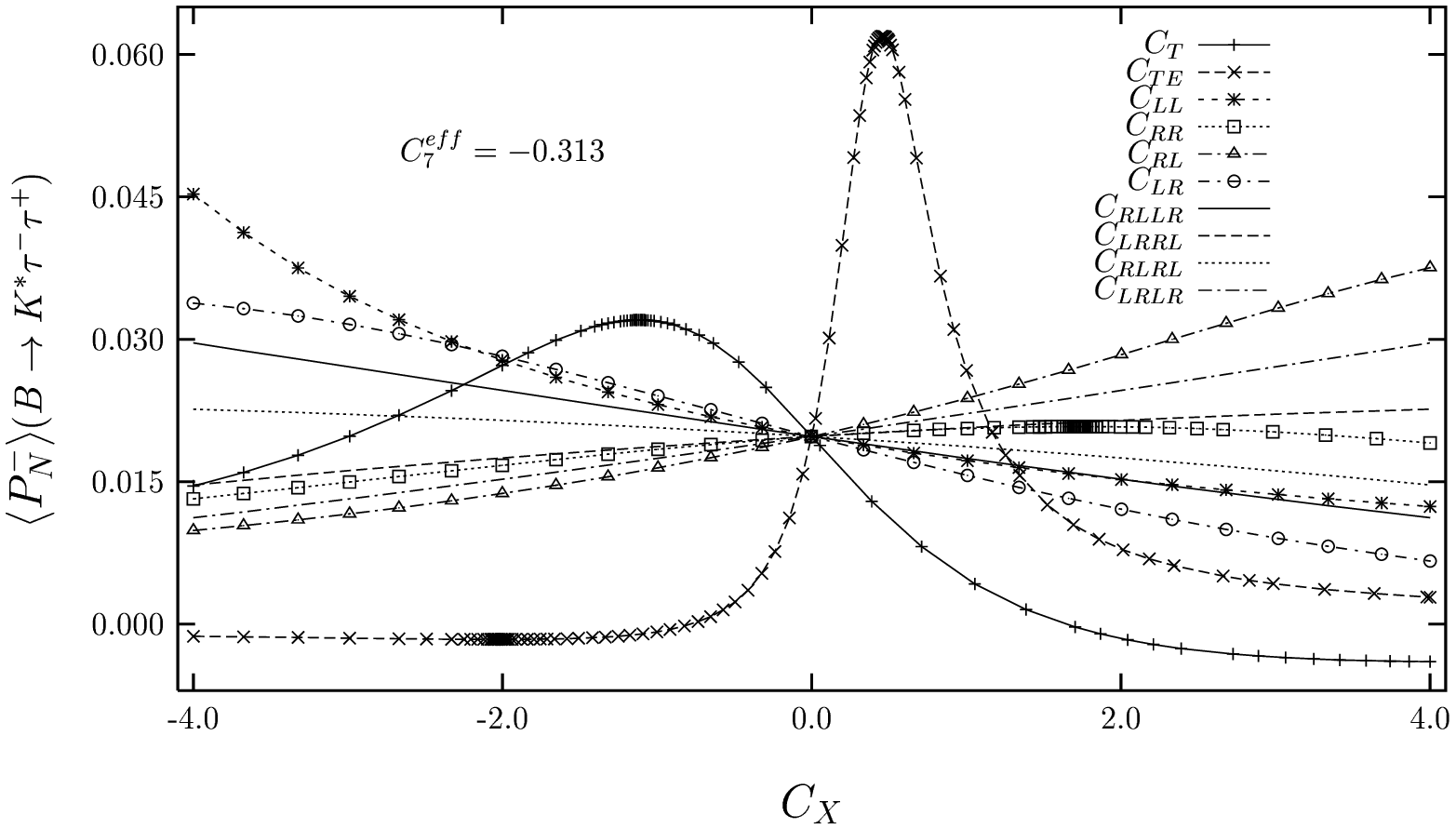}
\vskip 9. cm
\caption{}
\end{figure}

\begin{figure}
\vskip 1.5 cm
    \includegraphics{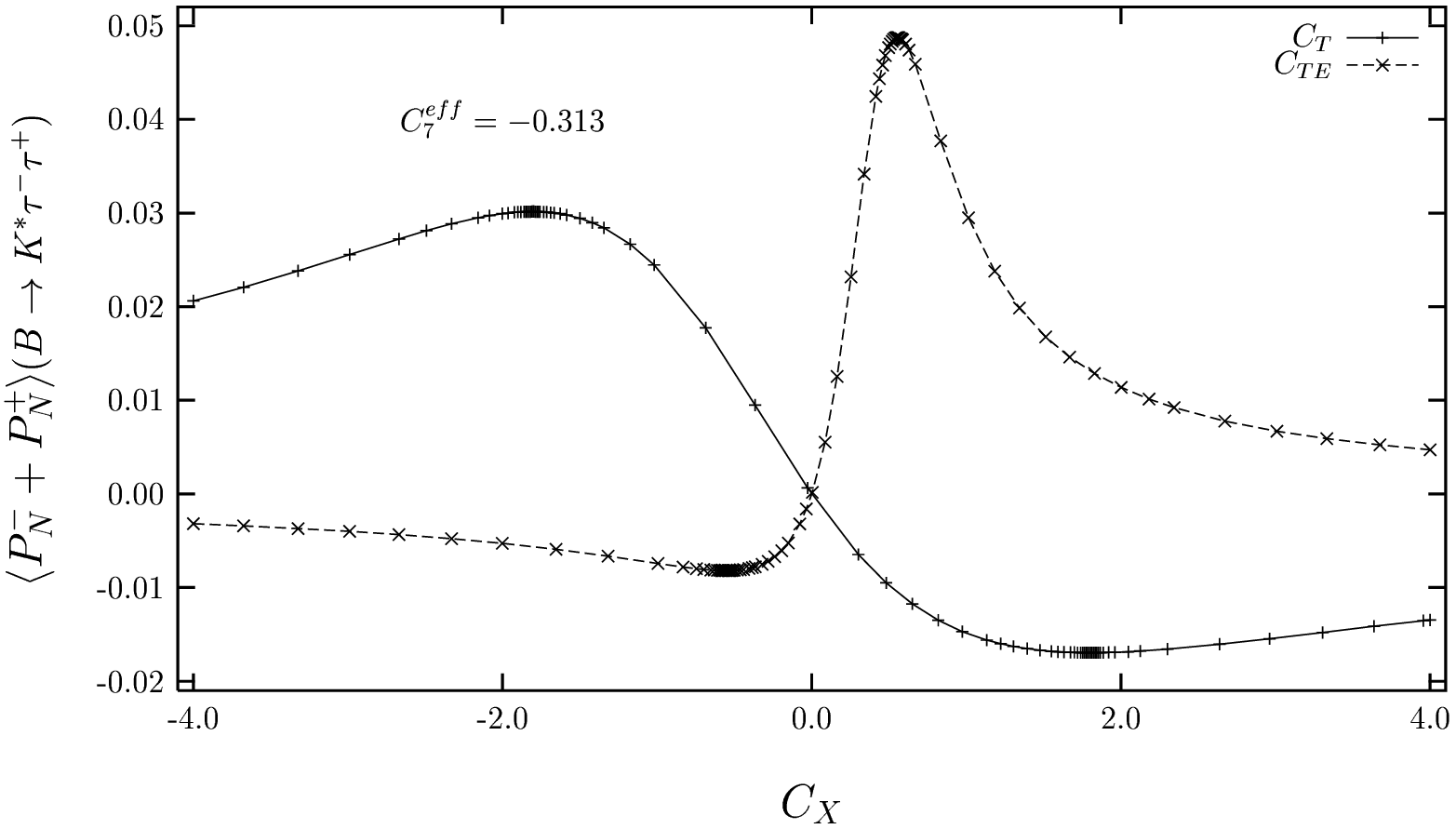}
\vskip 9. cm
\caption{}
\end{figure}

\begin{figure}
\vskip 1cm
    \includegraphics{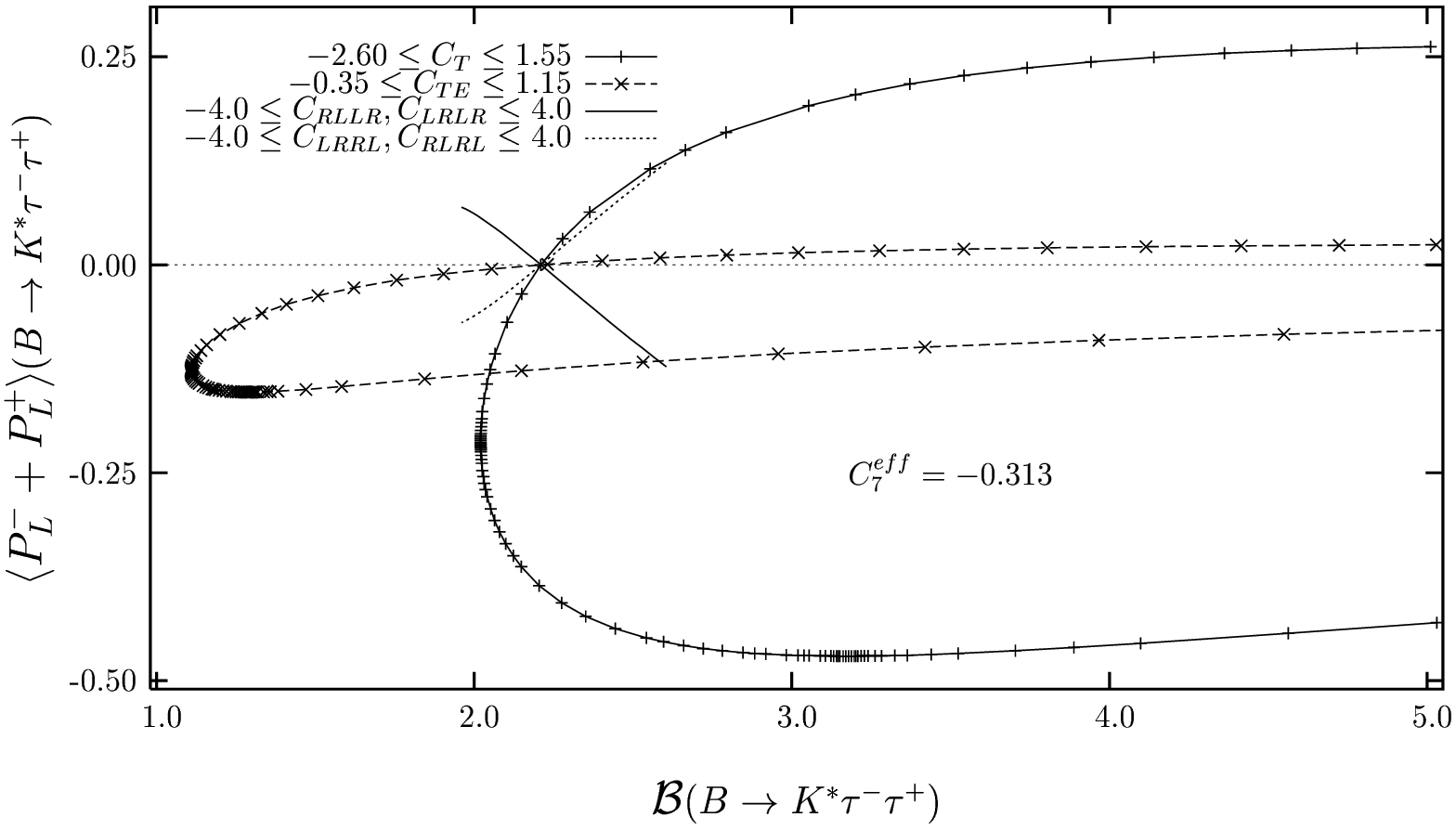}
\vskip 8.1cm
\caption{}
\end{figure}

\begin{figure}
\vskip 1.5 cm
    \includegraphics{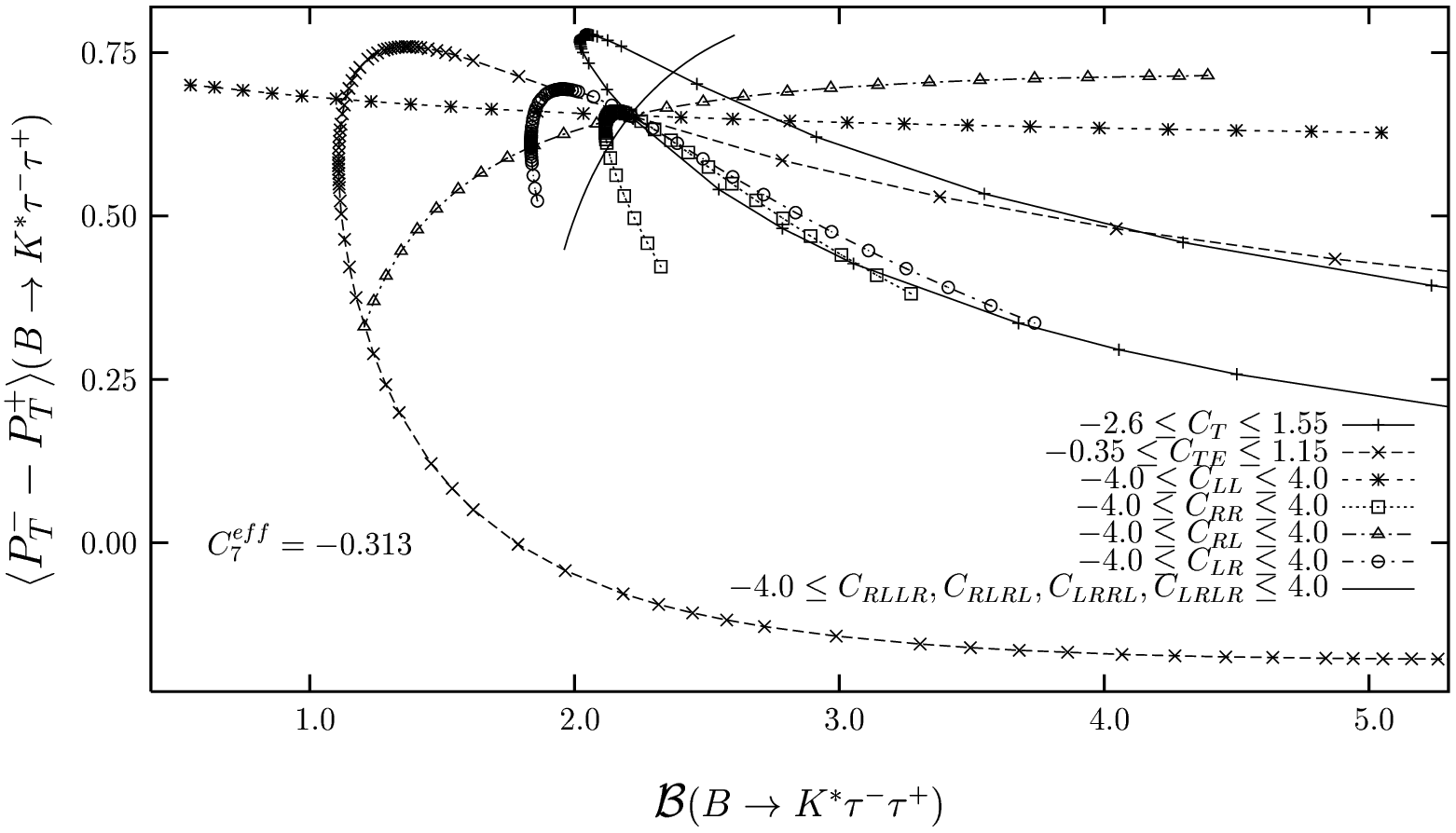}
\vskip 9. cm
\caption{}
\end{figure}

\begin{figure}
\vskip 1cm
    \includegraphics{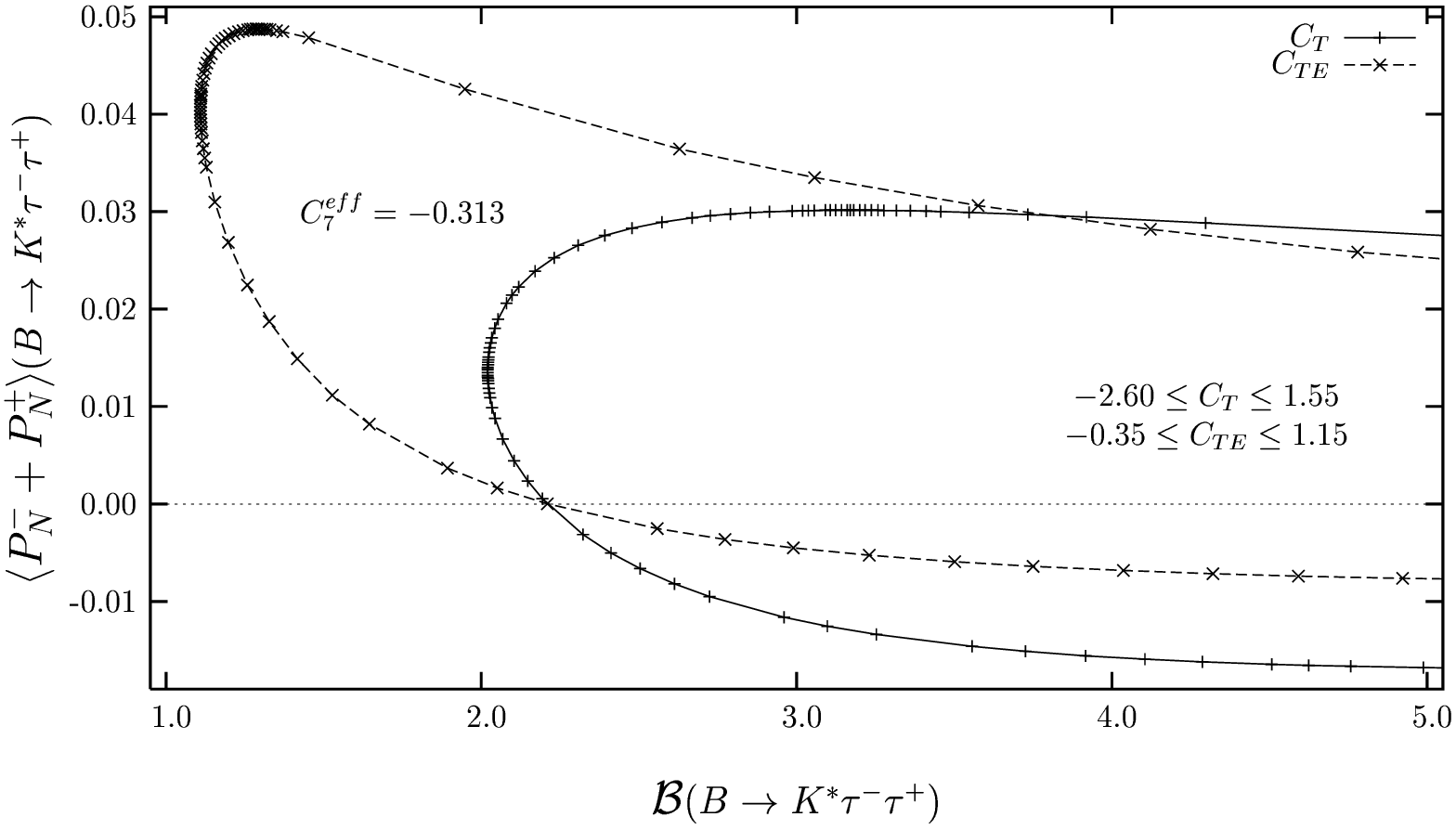}
\vskip 8.1cm
\caption{}
\end{figure}

\end{document}